\documentclass[11pt]{article}
\usepackage{amssymb,amsmath,amsthm,latexsym,mathrsfs,array,amssymb,amsmath,epsfig,sudoku,amssymb,amsbsy,array,color,amssymb,amsmath,epsfig,natbib,url,graphicx,soul,multirow}
\usepackage{float}
\usepackage{geometry,natbib,url}
\usepackage{comment}

\theoremstyle{Example}
\newtheorem{Example}{Example}[section]
\AtEndEnvironment{Example}{\null\hfill\qed}%
\newtheorem{Theorem}{Theorem}[section]
\newtheorem{Proposition}{Proposition}[section]
\makeatletter 
\newenvironment{Proof}[1][\proofname]{\par
	\normalfont
	\topsep6\p@\@plus6\p@ \trivlist
	\item[\hskip\labelsep\bfseries
	#1:]\ignorespaces
}{%
	\qed\endtrivlist
}
\makeatother

\newtheoremstyle{Example}{\topsep}{\topsep}%
{}
{}
{\bfseries}
{:}
{0.9mm}
{\thmname{#1}\thmnumber{ #2}\thmnote{(\it #3)}}

\theoremstyle{Example}
\AtEndEnvironment{Example}{\null\hfill\qed}%

\begin{document}

	\begin{center}
	{\bf\Large Credible Sets of Overall Ranking of Entities}
	\vskip 5pt
	Snigdhansu Chatterjee$^{a}$, Gauri Sankar Datta$^{b,c}$, Yiren Hou$^{d}$ and Abhyuday Mandal$^{b}$\footnote{Address for correspondence: Abhyuday Mandal, Department of Statistics, University of Georgia, Athens, GA, 30606 (amandal$@$stat.uga.edu).}
	
	$^{a}$University of Maryland, Baltimore County, Baltimore, MD\\
	$^{b}$Department of Statistics, University of Georgia, Athens, GA\\
	$^{c}$Center for Statistical Research and Methodology, U.S. Census Bureau, Suitland, MD \\
	$^{d}$Biostatistics, Yale School of Public Health, New Haven, CT
	
\end{center}

\vspace{.5in}

	


\begin{abstract}
Ranking, and inferences based on ranks,  are important problems in numerous contexts. This is especially true in small area statistics where there may be only a limited amount of directly observed data from each entity or small area, while precise and accurate estimates of best or worst performing entities are essential for fund allocation, planning and policymaking, stakeholder advocacy, evaluation of welfare programs,  and so on. Rankings based solely on point estimates, however, ignore uncertainty and may lead to inequities, particularly with small sample sizes. We propose novel Bayesian approaches to address this problem.  Our methods partition the parameter space into subsets with a posterior probability-driven partial ordering. Using a $(1-\alpha)$-coverage credible set, we obtain simultaneous inference for small area means and their ranks when $\alpha \in (0, 1)$; we also discuss the case $\alpha=0$. Our Bayesian algorithms substantially outperform state-of-the-art non-Bayesian alternatives while accommodating covariates and borrowing strength across small areas. We evaluate their accuracy and stability through applications and simulation studies. Finally, we develop a novel theoretical framework for ranking and inference under triangular arrays of Fay-Herriot models and establish probabilistic performance guarantees for the proposed Bayesian algorithms.
\end{abstract}

\noindent%
{\it Keywords:}  Fay-Herriot model; hierarchical Bayes; highest posterior density; small area estimation; triangular Fay-Herriot framework; unstructured Bayes
\vfill

\newpage

\section{Introduction}\label{intro}

International and national statistics offices (NSO) around the world routinely report on agriculture, employment, health, education, and several other aspects of human activities and well-being at a subpopulation or \textit{small area} level. Prominent examples include various statistics reported for the United Nations Sustainable Development Goals, the US Small Area Income and Poverty Estimation (SAIPE) program,  many welfare programs in the European Union (see \cite{heady2005eurarea}, for example), and poverty alleviation programs in many countries.  

In such small area contexts, estimation of subpopulation means of different variables is often the primary goal of NSOs. However, inference on the \textit{overall ranking of entities}, such as hospitals, animal herds, crop varietals, is equally critical in planning, policy making, ensuring fairness and advocacy related to such programs.
{For example, \cite{wright2026estimated} documented three noteworthy projects of the  U.S. Census Bureau that provide explicit estimated rankings. These include the American Community Survey (ACS) ranking tables, the Community Resilience Estimates (CRE) ranking tables and \textit{The Ranking Project} presented in \url{https://www.census.gov/topics/research/stat-research/ranking.html}.  This project provides novel theoretically animated visuals of estimated rankings of the 51 states on the same 89 topics covered by the ACS Ranking Tables for the years 2018-19, 2021-24. Additional visualizations to aid stakeholders with their planning and advocacy is available here: \url{https://www.census.gov/csrm/rankings/ranking-ranges.html}. Additional applications on ranking of survey-derived products in the U. S. Census Bureau may be found in \cite{bell2003,gomez2010bayesian, wright2014ranking}. Ranking  and appreciation of related uncertainties is supremely relevant for other  small area applications also: for identifying vulnerable populations and neighborhoods \citep{willyard2025developing}, agriculture \citep{bijari2025method}, transportation \citep{aminghafouri2024joint},  public health issues like cancer prevention and obesity \citep{moss2017comparing, waldrop2017ranking, al2022simultaneous}  to modern day artificial intelligence and data science applications \cite{chatzi2024prediction, neuhof2024confident, chen2025ranking, liao2026distribution, rising2021uncertainty}. 
}
\cite{LairdLouis1989} noted that ranking draws attention to unusually  high or low performing subpopulations and provides  investigators and policy makers useful tools to establish priorities. Since available resources are limited,  authorities need to identify the most underprivileged or impoverished subpopulations to provide assistance. 
Similar challenges arise in other types of applications where there is limited amount of directly observed data on 
subpopulations or subdomains of interest; for example, in precision medicine and various biostatistical problems, precision agriculture, or urban engineering problems.  Hereafter, we use the generic term ``small area'' or ``entity'' to imply such subdomains where there is paucity of directly observed data. In this paper, we address the problem of ranking $m$ such entities, indexed by $i = 1, \ldots, m$, based on a common characteristic $\boldsymbol{\theta} = (\theta_{1}, \ldots, \theta_{m})$, which is reflected in a measurable quantity $\mathbf{Y} = (Y_1,\cdots, Y_m)$.

\begin{Example} [Ranking US states by commute times]
\label{Exmp:1}
As an illustrative example that serves to both motivate and contextualize the proposals of this paper as well as set notations, consider the problem of ranking the  states of the USA and Washington, DC, based on the average commuting times.  Let $\theta_{i}$ be the unknown true average commuting time in state $i \in \{1, \cdots, m (=51) \}$.  Define the \textit{true rank vector} $\check{\mathbf{r}} = (\check r_{1}, \ldots, \check r_{m})$, where 
	\begin{equation}
	\label{truerank}
	\check r_i \equiv \check r_i(\boldsymbol{\theta}) = \sum_{j=1}^m I(\theta_j \le \theta_i) = 1 +  \sum_{j: j \ne i} I(\theta_j \le \theta_i),~
	\mbox{for } i=1,\cdots, m.
	\end{equation}
Suppose limited data $\mathbf{Y} = (Y_1,\cdots, Y_m)$ are available, based on which we make predictions $\hat{\boldsymbol{\theta}} = (\hat{\theta}_{1}, \ldots, \hat{\theta}_{m})$ on $\boldsymbol{\theta} \in \boldsymbol{\Theta}_{m} \subseteq \mathbb{R}^{m}$. It should be emphasized that in keeping with the reality of small area estimation (SAE) problems, $\hat{\boldsymbol{\theta}}$ is not an accurate or consistent estimate of $\boldsymbol{\theta}$; indeed, both are often $O_{P} (1)$ random variables sharing some common features. 

Using $\hat{\boldsymbol{\theta}}$ we may obtain estimator $\hat{\mathbf{r}} = (\hat{r}_{1}, \ldots, \hat{r}_{m})$ of $\check{\mathbf{r}}$ as below:
	\begin{equation}
	\label{obsdrank}
	\hat{r}_{i} = \sum_{j=1}^m I(\hat{\theta}_j \le \hat{\theta}_i) = 1 +  \sum_{j: j \ne i} I(\hat{\theta}_j \le \hat{\theta}_i),~
	\mbox{for } i=1,\cdots, m.
	\end{equation}
Obviously, sampling errors and chance variability of the $Y_i$'s lead to errors in the $\hat{r}_{i}$'s, consequently quantifying and measuring  their uncertainty is important. To quantify and account for uncertainty in $\hat{\mathbf{r}}$, we need to study the \textit{joint distribution} of $\hat{\boldsymbol{\theta}}$. 
\end{Example}

Compared to many other aspects of small area statistics, the problem of \textit{ranking of small areas} has received relatively less attention; we discuss the relevant literature in Section~\ref{Sec:Review}. Note that several of the existing methodological approaches are on producing  \textit{a single rank vector}, that is, a permutation of the indices $\{1, 2, \ldots, m \}$, as an output.  Since in small area contexts there is no notion of consistency of $\boldsymbol{\theta}$-estimation and associated quantities, a single vector of ranks is not particularly useful for policymakers or stakeholders.

To address this gap, in a recent pioneering paper, \cite{KWW2020} considered the problem of producing candidate predictors for the rank vector $\check{\mathbf{r}}$ with some uncertainty quantification, using a clever frequentist approach, see Section~\ref{Sec:KWW} for a review. In the rest of this paper we use the abbreviation KWW to refer to this paper or the methodology proposed by these authors. Their technique allows any entity or small area unit to hold multiple ranks, and any  rank to  potentially apply to multiple units, which is a major innovation over preceding approaches. \cite{KWW2020} cautioned against naively using point estimates to  rank subpopulations and thus ignoring the related uncertainties, while noting that the published ranking of various small area entities  based on  experimental, observational or  survey data  (to quote) ``\textit{are usually released without direct statistical statements of uncertainty on estimated overall rankings}''.

However, multiple major innovations are needed on the KWW approach for any ranking procedure to be truly useful in small area problems and related topics. First, the above frequentist method of KWW provides no opportunity to utilize covariates  related to the $\theta_i$'s, or to borrow strengths across small areas. Second, owing to its construction and the lack of a principled way to borrow strength, the resulting confidence intervals (sets) are too wide (voluminous). Although KWW provides multiple plausible rank vectors, there is no scheme or guidance to suggest which of these vectors may be closer to reality compared to others; in other words, all reported rank vectors are equal \textit{per se}. There is also no existing theoretical basis for the various methodological choices implied in KWW.

In this paper, we address all these challenges, using the Bayesian paradigm as our mainstay. We introduce two  Bayesian models by specifying suitable prior distributions for the unknown $\boldsymbol{\theta}$ on $\boldsymbol{\Theta}_{m}$.  The first  prior, stated in  equation~\eqref{unif-th-pr} below, is noninformative and flat over $\boldsymbol{\Theta}_{m}$.  For this prior, one possible choice of a credible set for $\boldsymbol{\theta}$  exactly matches the  confidence set developed by \cite{KWW2020}, and hence allows us to compare the proposals of this paper with that of KWW.  The second prior, stated in Section~\ref{HBayes}, leads to the widely used hierarchical Bayes (HB) and empirical Bayes (EB) models  studied by \cite{fay1979estimates}, \cite{berger1985book}, \cite{AitkinLongford1986}, \cite{BergerDeely1988}, \cite{LairdLouis1989}, \cite{morris1996hierarchical} and many others.

With either of these priors, we first obtain {a size $(1- \alpha)$-{\it credible set}}  for $\boldsymbol{\theta}$ (generically denoted by $\mathcal{S} \subset \boldsymbol{\Theta}_{m}$, with qualifying subscripts as needed), for a user specified value of $\alpha \in [0, 1)$. We obtain rank vectors from each of these candidate values, and thus we obtain the corresponding {empirical credible set} of the rank parameter $\check{\mathbf{r}}$, generically denoted by $\mathcal{R} \subset \sigma(\{1, 2, \ldots m \})$ (the set of all permutations of $\{1, 2, \ldots m \}$), again with appropriate subscripts as required. Typically, there will be many  candidate $\boldsymbol{\theta}$ within the $\mathcal{S}$, and in principle uncountably infinite number of such choices exist in many problems. Many of these candidates will produce the same rank vector, since \eqref{obsdrank} is a many-to-one function.  If we order the candidate ranks in $\mathcal{R}$ ensuring their frequencies are non-decreasing (or non-increasing), we can obtain a \textit{partially ordered set} (poset, in short), which in an abuse of notation we will also call $\mathcal{R}$ generically. This partial ordering induces a corresponding partial ordering in $\mathcal{S}$, and in the rest of the paper we will also consider $\mathcal{S}$ to be a partially ordered set. In the various applications and simulation studies that we conducted, the partially ordered set $\mathcal{R}$ produced by either of the two priors, and two different techniques to obtain credible sets $\mathcal{S}$, resulted in significantly better quality and more easily interpretable results compared to KWW.

For example, an analysis using 9 states with the lowest commuting times in Example~\ref{Exmp:1} showed that the 90\% confidence set obtained by the KWW method has more than $40,000$ plausible vectors compared to only about $400$ rank vectors from a 90\% empirical credible set  $\mathcal{S}$.  The fact that our Bayesian approach can elicit a relative weight associated with each rank vector and thus present an \textit{ordering} of the credible rank vectors is exceptionally useful for planners, policymakers, and stakeholders.  As an illustration based on Example~\ref{Exmp:1}, if we consider the four states SD, ND, WY and  NE that have the least commuting times,  a 90\% joint credible set of the rank vector is $\{4321, 4312, 3421, 3412, 4231, 4213, 4132, 4123\}$, with associated weights 0.2935, 0.2722, 0.2082, 0.1835, 0.0232, 0.0140, 0.0032 and 0.0021 (see supplementary Table~\ref{Table:Supple_4States_New}). In contrast, the KWW approach obtains 14 vectors as an unordered set of ranks. The difference in the sizes of a Bayesian credible set and a KWW-based confidence set with identical coverage increases rapidly with an increasing $m$, see supplementary Table~\ref{Tab:Supple_NumerOfRankVectors}.  Similar comments can be made for EB-based approaches resulting from our computations, but since such details bring no additional insight, we omit these  to contain the length of this paper.

Apart from the novel Bayesian methodological developments that address all the practical challenges associated with ranking small areas and using those for official purposes, we address the challenge of developing a framework for understanding the theoretical properties of predictors of ${\boldsymbol{\theta}}$ like $\hat{\boldsymbol{\theta}}$ or rank-predictors like $\hat{\mathbf{r}}$ for $\check{\mathbf{r}}$. Note that $\hat{\boldsymbol{\theta}}$ is not a consistent estimator of ${\boldsymbol{\theta}}$, that is, $\hat{\boldsymbol{\theta}} - {\boldsymbol{\theta}} = O_{P} (1)$ irrespective of the data sample size $m$. However, even under such a strong restriction, could there be notions of convergence associated with the posterior distributions of ${\boldsymbol{\theta}}$ or $\check{\mathbf{r}}$? In Section~\ref{Sec:Theory} we establish that  indeed there can be such convergence notions under very reasonable conditions, using recent developments on comparing probability measures on spaces of different dimensions. For this novel theoretical development, we introduce and develop the \textit{triangular Fay-Herriot} ($\Delta$-FH) framework, and show that for posterior distributions of ${\boldsymbol{\theta}}$ or $\check{\mathbf{r}}$, Cauchy convergence type results hold almost surely. Our theoretical results have wider implications for small area studies beyond establishing properties of EB or HB small area ranking problem, and such subsequent developments will be reported in depth in a future study.

The rest of the paper is organized as follows. In Section~\ref{Sec:Review} we review existing techniques for ranking small areas and entities, with Section~\ref{Sec:KWW} containing  a detailed review of \cite{KWW2020}, and Section~\ref{Sec:MRSW} containing a review of a  variation of the KWW algorithm studied in \cite{Mogstad2024} (MRSW hereafter). Then in Section~\ref{Sec:Bayesian} we expand on the details of our Bayesian proposal. We discuss two choices of priors: a flat, noninformative one and another that is traditionally used in small area statistics. We also study two different ways of creating credible sets: an orthotope (hyperrectangle) and a hyperellipse, to create partially ordered sets of $\boldsymbol{\theta}$ and their corresponding ranks. Details of the algorithm are presented, along with techniques for presentation of the ranking results. While our studies encompass four possible Bayesian solutions,  the basic principles we lay down are applicable to any other reasonable choice of prior or technique for constructing credible sets. In Section~\ref{Sec:Interpretation} we discuss in detail the interpretation of our Bayesian results. The special case corresponding to $\alpha = 0$ (and thus using a $100\%$ credible set), which is appealing but has limited functionality in real life applications, is discussed in detail in Section~\ref{Sec:DirectRanking}.  Section~\ref{Sec:Simulation} contains a simulation experiment to illustrate the advantages and limitations of our proposed Bayesian approaches and their performance relative to KWW and MRSW. Section~\ref{kww-commuting} contains a detailed discussion on the example on commuting times in US states that is briefly described in Example~\ref{Exmp:1}. In Section~\ref{medianincome} we discuss another important small area problem of ranking US states using the median income, where some ground truths are known. Section~\ref{Sec:Theory} presents the $\Delta$-FH framework and associated theoretical and technical developments. Section~\ref{Summary} provides a summary and a few concluding remarks. Some supporting results and technical developments  are presented in the Supplement.

\section{Ranking of small areas: a review}
\label{Sec:Review}

 Some of the very early related work on ranking and selection are by \cite{Bechhofer1954} and \cite{gupta1989decision}. Using the Bayesian technique of  ranking several binomial populations by \cite{BlandBratcher1968}, \cite{GovindarajuluHarvey1974} presented a unified  treatment of the ranking and selection problem. \cite{GoelRubin1977} proposed a Bayesian approach on selecting a subset containing the best subpopulation. \cite{LairdLouis1989} used an EB approach to monitoring regions suspected of elevated disease rates for residents or health service providers suspected of elevated  service failure rates for patients, which requires attention to mitigate risks. In fact, \cite{morris1996hierarchical} highlighted the utility of the Bayesian approach in ranking of hospitals. An early application of modeling of  $\boldsymbol{\theta}$ through regression on relevant covariates is due to \cite{EfronMorris1975}. Ranking thirty-six cities of El Salvador based on toxoplasmosis rates, they indicated the pitfall of ranking using only observed rate estimates but ignoring their variability.       
To estimate $\check{\mathbf{r}}$, \cite{AitkinLongford1986} considered setups where in one case they assumed that the $\theta_i$'s are non-random, and in the second case they modeled $\theta_i$'s using a probability distribution, possibly depending on parameters ($\boldsymbol{\psi}$, say). The latter case uses a hierarchical modeling for $\mathbf{Y}$ and $\boldsymbol{\theta}$,  where $\check{\mathbf{r}}$ is estimated by computing first an estimator of $\boldsymbol{\psi}$ and a predictor of  $\boldsymbol{\theta}$. 
 \cite{BergerDeely1988} used a prior for $\boldsymbol{\psi}$ and proposed a HB approach. They tested a null hypothesis ($H_0$) for equality of subpopulation means, and conditional on $H_0$ is false, they estimated probability of $\theta_i$ being the largest of the $m$. They estimated the ranks of the $\theta_i$'s based on the ranks of these probabilities. Although \cite{AitkinLongford1986}, \cite{BergerDeely1988}, \cite{LairdLouis1989}, \cite{GoldsteinSpiegelhalter1996JRSSA},  \cite{morris1996hierarchical} and \cite{ShenLouis1998} considered EB/HB estimation of individual ranks, none considered  set estimation of  $\check{\mathbf{r}}$. Similarly, \cite{bell2003} used the posterior distribution of $\boldsymbol{\theta}$ corresponding to a flat prior to obtain marginal posterior distributions of individual ranks, but did no inference or credible sets.   Consequently, all these studies lacked uncertainty quantification for the elicited rank vector, and are less useful for policy makers and stakeholders.

\subsection{The KWW approach}\label{Sec:KWW}

Let $\mathbf{D} = diag(D_1,\cdots, D_m)$ be known variances. 
\cite{KWW2020} assumed  the model 
	\begin{equation}
	\label{sampling}
Y_i | \boldsymbol{\theta}\; \stackrel{ind}\sim \; N(\theta_i, D_i), ~ i =1,\cdots, m.
	\end{equation}
{Using independent $Y_i$'s, for} $\alpha \in (0, 1)$,  a $(1-\alpha)$-level joint confidence set $\mathcal{S}_{KWW, \alpha}$ for $\boldsymbol{\theta}$ is constructed as the Cartesian product of the 
confidence intervals $I_i = [Y_i - z_{1-\frac\gamma 2}\sqrt{D_i}, Y_i - z_{\frac \gamma 2}\sqrt{D_i}] \equiv [L_i, U_i]$, $ i=1,\cdots, m,$. Here, $z_q$ is the standard normal $q$-th quantile and $\gamma = 1 - (1-\alpha)^{1/m}$. 
Then for each $i \in \{1, 2, \cdots, m\}= \mathbb{N}_{m}$ and $C_i = \mathbb{N}_{m}\setminus \{i\}$, KWW defined three subsets 
$	\Lambda_{Li} = \{ j \in C_i: U_j \le L_i\}$,
	$\Lambda_{Ri} = \{ j \in C_i: U_i \le L_j\}$,
	$\Lambda_{Oi} =     C_i \setminus \{\Lambda_{Li} \cup  \Lambda_{Ri}\}$. 
Using these subsets of $\mathbb{N}_{m}$, \cite{KWW2020} showed that the set of rank vectors 
	\begin{equation}
	\label{KWWrankset} 
	\mathcal{R}_{KWW, \alpha} = 
	\{(r_1,\cdots, r_m):   r_i \in \{|\Lambda_{Li}| +1, 
    \cdots, |\Lambda_{Li}| +1 +|\Lambda_{Oi}|\} ~\mbox{for}~ i=1,\cdots, m \},
	\end{equation}
is a joint confidence set for $\check{\mathbf{r}}$ having a joint coverage probability at least $(1-\alpha)$. KWW have briefly remarked that a Bayesian version 
 of their frequentist solution is possible. Our approach is  different from what \cite{KWW2020} alluded to in that remark, and requires the use of the Bayesian paradigm and  Bayesian computational techniques. 
 
\subsection{A modification of the KWW approach}	
\label{Sec:MRSW}

 The recent paper by \cite{Mogstad2024} (MRSW) extended the KWW work by considering joint confidence intervals of all pairwise differences of the form $(\theta_i -\theta_j)$ (see also \cite{wright2026estimated}). 
Define $\mathbb{S}_{m,m} = \{(i,j) \in \mathbb{N}_{m} \times \mathbb{N}_{m}: i \ne j\}$. For $(i,j) \in \mathbb{S}_{m,m}$, define $\eta_{i,j} = \theta_i -\theta_j$. MRSW used the collection of pivotal quantities $W_{i,j} = (Y_i-Y_j - \eta_{i,j})/\sqrt{D_i+D_j}$ to construct $1-\gamma$ confidence intervals for the $\eta_{i,j}$'s, for  an appropriate $\gamma$ such that  the joint confidence level is $1-\alpha$. For any $i \in \mathbb{N}_{m}$, let $F_i^-$ ($F_i^+$) be the number of confidence intervals of $\eta_{i,j}$ that are entirely left (right) of zero. MRSW showed that
\begin{equation}
\mathcal{R}_{MRSW, \alpha} = 
\bigl\{(r_1,\cdots, r_m):   r_i \in \{F_i^- +1,  
\label{MRSWrankset}
    \cdots, m -F_i^+\} ~\mbox{for}~ i=1,\cdots, m \bigr\},
\end{equation}
is a joint confidence set for $\check{\mathbf{r}}$ with a 
coverage probability at least $(1-\alpha)$.
By $|\Lambda_{Li}| +|\Lambda_{Oi}| + 1
 = m-|\Lambda_{Ri}|$, from (\ref{KWWrankset}) and (\ref{MRSWrankset})
 the similarity of the KWW and MRSW solutions reveals.

\section{The proposed Bayesian framework}	
\label{Sec:Bayesian}

\subsection{An unstructured Bayesian (UB) model}\label{unstrBayes}
	
We propose augmenting the sampling model in \eqref{sampling} using the prior for $\boldsymbol{\theta} \in \boldsymbol{\Theta}_{m}$ 
	\begin{equation}
	\label{unif-th-pr}
	\pi(\theta_1,\cdots, \theta_m) = 1 ~
	\mbox{for} ~ -\infty <\theta_1 <\infty,\cdots, -\infty <\theta_m <\infty,
	\end{equation}
which is an improper and noninformative prior. 
The posterior distribution of $\boldsymbol{\theta}$ given $\mathbf{Y} = \mathbf{y}$ is $N( \mathbf{y}, \mathbf{D})$. Note that the highest posterior density (HPD) $1-\gamma$ credible interval for $\theta_i$ is identical to confidence interval $I_i$ for $\theta_i$ that was the cornerstone of the \cite{KWW2020} approach, and  the Cartesian product $\prod_{j = 1}^{m} I_j $ is a joint credible set for $\boldsymbol{\theta}$, which is identical to the joint confidence set constructed by \cite{KWW2020}. 

While the UB model helps reproduce the KWW confidence set of $\boldsymbol{\theta}$, it fails to address the impact of one or more large  $D_i$ on the credible set of  $\check{\mathbf{r}}$. If $y_i > y_j$, then, irrespective of $D_i, D_j$,
$P[\theta_i > \theta_j| \mathbf{y}] >1/2$. 
\cite{BergerDeely1988} and \cite{morris1996hierarchical} justifiably criticized this result as unsatisfactory.

\subsection{A hierarchical  Bayesian (HB) model}\label{HBayes}

To address the shortcomings of the UB model, we also study the use of the prior
\begin{align}
		\theta_i|\boldsymbol{\beta}, A \stackrel{ind}\sim N(\mathbf{x}_i^T\boldsymbol{\beta}, A), \text{ for } i=1,\cdots, m
		\label{eq:FH_Prior}
\end{align}		
to pair with \eqref{sampling}. Here the regression coefficient $\boldsymbol{\beta} \in \mathbb{R}^{p}$ and variance $A$ are  hyperparameters for the linking model. The known vector $\mathbf{x}_i$ is the covariate for $Y_i$. This model was used by \cite{fay1979estimates} in small area estimation context; cf.  
the \textit{Fay-Herriot model} (e.g., p. 77 of \cite{rao2015small}). It is extensively used in many governmental, international and official agencies across the world for measuring poverty, healthcare-related indices and numerous other human well-being-related official data products. A multitude of other uses for this model may be found in  \cite{EfronMorris1975}, \cite{berger1985book}, \cite{BergerDeely1988} and \cite{morris1996hierarchical} and elsewhere.
The above Fay-Herriot model can be used easily to obtain the posterior distribution of $\boldsymbol{\theta}$, conditional on  $\mathbf{Y} = \mathbf{y}$ and hyperparameters $\boldsymbol{\beta}$ and $A$, as well as the marginal distribution of $\mathbf{Y}$, which depends on the hyperparameters. 
In our data applications presented below involving the  HB framework, we use the improper prior
$\pi(\boldsymbol{\beta},A) = {(\bar{D} +A)^{-\frac{1}{2}}}$,  where $\bar D = \sum_{i=1}^m D_i/m$. The Bayesian computations are insensitive to reasonable choice of priors, and, in particular, the above prior(s) result(s) in a fast and efficient MCMC algorithm.


\subsection{A Cartesian  empirical credible set construction}\label{Cartesian}

Using either of the priors above, we generate a large  sample of size $S$ from the posterior distribution of  $\boldsymbol{\theta}$. Generically, we denote these sample values by the set $\boldsymbol{\Theta}^{(S)} = \{\boldsymbol{\theta}^{(s)} \in \boldsymbol{\Theta}_{m}: s = 1, 2, \ldots, S\}$. For the UB model of Section~\ref{unstrBayes},  $\boldsymbol{\theta}^{(s)}$ is drawn from $N(\mathbf{y}, \mathbf{D})$. For the HB model of Section~\ref{HBayes}, the set $\boldsymbol{\Theta}^{(S)}$ is drawn by MCMC. See Section~\ref{full-condnls} in the Supplement for details. We use the samples in $\boldsymbol{\Theta}^{(S)}$ for the  construction of an empirical credible set $\mathcal{S}$. 

When $\mathcal{S}$ is an orthotope,  we obtain the empirical one-dimensional credible set for  $\theta_j$ for  each $j \in \{ 1, 2, \ldots, m \}$;
we consider the $j$-th element from the vectors in $\boldsymbol{\Theta}^{(S)}$, namely $\theta_j^{(1)}, \cdots, \theta_j^{(S)}$. For any $\delta \in (0, 1)$, the  $(\delta/2)$th and $(1 -\delta/2)$th sample quantiles of these, say $\theta_j^{[\delta/2]}$ and  $\theta_j^{[1 - \delta/2]}$, form the interval $S_{j, \delta} = (\theta_j^{[\delta/2]}, \theta_j^{[1 - \delta/2]})$.
Now define their Cartesian product $ \mathcal{S}_{*, C, \alpha} = \prod_{j=1}^m S_{j, \delta}$ where the subscript $``*"$ is either UB or HB depending on whether the prior is as given in \eqref{unif-th-pr} or in \eqref{eq:FH_Prior}. Here, $\delta$ is chosen such that the number of elements of $\boldsymbol{\Theta}^{(S)}$ in  $\mathcal{S}_{*, C, \alpha}$ is the integer closest to $ S(1-\alpha)$. We use the additional 
subscripts of $C$  and $\alpha$ to emphasize that the set is a Cartesian product and depends on $\alpha$.

\subsection{Construction of elliptical  empirical credible sets}\label{elliptical}

An optimal method to {create a credible set} $\mathcal{S}$ from a posterior is to use the highest posterior density (HPD) method. An HPD credible set is of the form $\{\boldsymbol{\theta}: \pi (\boldsymbol{\theta}| \mathbf{y}) \ge t \}$, where  $t$ is chosen such that the posterior probability of this set is $(1-\alpha)$. Among all $(1-\alpha)$ credible sets, the HPD credible set has the lowest volume. 
For the UB model, $\pi (\boldsymbol{\theta}| \mathbf{y})$ is a decreasing function of  the Mahalanobis distance 
$M(\boldsymbol{\theta}, \mathbf{y}, \mathbf{D}) = (\mathbf{y}-\boldsymbol{\theta})^T \mathbf{D}^{-1} (\mathbf{y} -\boldsymbol{\theta})$.  We construct the empirical HPD credible set  $ \mathcal{S}_{UB, E, \alpha} = \{\boldsymbol{\theta} \in \boldsymbol{\Theta}^{(S)}: M(\boldsymbol{\theta}, \mathbf{y}, \mathbf{D}) \le c \}$,  where $c$ is chosen such that the empirical posterior probability of $\mathcal{S}_{UB, E, \alpha}$ is at least $1 - \alpha$. Here, the subscript ``$E$'' denotes that this empirical credible set is based on a hyperellipse. 

For the HB model,  for large $m$, $\pi_{HB} (\boldsymbol{\theta} | \mathbf{y})$ is {\it approximately} multivariate normal. This can be intuitively understood by noting that the conditional posterior distribution of $\boldsymbol{\theta}$ given $A$ is  multivariate normal and applying the Laplace approximation and consistency of the posterior mode of $A$; we omit the details. Thus, an HPD credible set for $\boldsymbol{\theta}$ will be approximately elliptic, defined by the Mahalanobis distance based on $\hat{\boldsymbol{\theta}}_{HB} = \frac 1S \sum_{s=1}^S \boldsymbol{\theta}^{(s)}$,  $V_{HB} = \frac 1S \sum_{s=1}^S \{\boldsymbol{\theta}^{(s)} - \hat{\boldsymbol{\theta}}_{HB}\}\{\boldsymbol{\theta}^{(s)} - \hat{\boldsymbol{\theta}}_{HB}\}^T$.
Similar to the UB case, the empirical credible set is $ \mathcal{S}_{HB, E, \alpha} = \{\boldsymbol{\theta}^{(s)} \in \boldsymbol{\Theta}^{(S)}: M(\boldsymbol{\theta}^{(s)}, \hat{\boldsymbol{\theta}}_{HB}, V_{HB}) \le c \}$ for  suitably chosen $c > 0$.

Note that in a likelihood-based approach one can construct an elliptic-type confidence set by inverting a likelihood-based acceptance region. However, due to the {\it interdependence} of the components of $\boldsymbol{\theta}$, it would not be possible to line these up to determine their rankings, a major hurdle in obtaining an optimal 
confidence set for $\check{\mathbf{r}}$.  In our Bayesian setup the shape of a credible set poses no challenge, as we obtain the rank vector of each $\boldsymbol{\theta}$ within a credible set directly. This  is a major advantage for the proposed Bayesian approach.

	

\subsection{Bayesian rank posets}\label{jt-cred-overall-rank}

Below, we use the notation $\mathcal{S}_{Bayes, \alpha}$ to denote any one of the four  empirical credible sets $ \mathcal{S}_{UB, C, \alpha}$, $ \mathcal{S}_{HB, C, \alpha}$, $ \mathcal{S}_{UB, E, \alpha}$, or $ \mathcal{S}_{HB, E, \alpha}$ constructed above.  For each element $\boldsymbol{\theta}^{(s_j)}$ in $\mathcal{S}_{Bayes, \alpha}$, we use equation~\eqref{obsdrank} to obtain a candidate rank vector $\mathbf{R}^{(s_{j})}$, which are collected in the set $\mathcal{R}_{Bayes, \alpha}$.  In addition, we also have the relative weights of the distinct elements of the set $\mathcal{R}_{Bayes, \alpha}$.  Arranging these in a non-decreasing order, we henceforth consider  $\mathcal{R}_{Bayes, \alpha}$ as \textit{ a partially ordered set} (poset) of rank vectors.  This partial ordering can be easily extended to $\mathcal{S}_{Bayes, \alpha}$ in  a natural way, so we consider those sets to be also posets. Also, the empirical credible sets can be extended to credible sets by taking the convex hull of $\mathcal{S}_{Bayes, \alpha}$, which would also be posets. Note however, that the KWW-based set $\mathcal{R}_{KWW, \alpha}$ from \eqref{KWWrankset} remains a set of unordered vectors, since there is no viable way of constructing an ordering of the elements of this set. On the other hand, from any of the the four choices for $\mathcal{R}_{Bayes, \alpha}$ we can easily obtain the marginal empirical frequency of any subvector of ranks, or the frequencies of the ranks for any subset of small areas. 

\begin{table}[h]\caption{Matrix $\mathbf{R}_{HB, E}\times 100$ for the 9 states from Example~\ref{Exmp:1} that have the lowest commuting times ($\mathbf{Y}$), using MCMC from the HB approach and elliptical credible set of 90\% coverage.}\label{tab:HB.HPD.9States_251021}
\begin{center}
{\scriptsize
\begin{tabular}{|c|rrrrrrrrr|}
      \hline
     & \multicolumn{9}{c|}{States} \\
Rank & ID & KS & IA & AK & MT & NE & WY & ND & SD \\
  \hline
  1 &      &       &      &      &       &       &  0.03 & 42.61 & 57.36 \\
  2 &      &       &      &      &  0.02 &  0.01 &  0.41 & 57.03 & 42.53 \\
  3 &      &       &      & 5.36 & 19.12 & 36.22 & 38.84 &  0.36 &  0.11 \\
  4 &      &       &      & 11.03& 28.65 & 39.91 & 20.40 &  0.01 &       \\
  5 &      &  0.15 & 0.14 & 25.83& 31.93 & 19.94 & 22.01 &       &       \\
  6 &      &  1.93 & 4.50 & 53.91& 19.72 &  3.92 & 16.01 &       &       \\
  7 &      & 27.85 & 67.47& 2.84 &  0.44 &       &  1.40 &       &       \\
  8 & 0.03 & 70.03 & 27.88& 1.03 &  0.12 &       &  0.90 &       &       \\
  9 & 99.97&  0.03 &      &      &       &       &       &       &       \\
   \hline
\end{tabular}
}
\end{center}
\end{table}

 For clarity in presentation to stakeholders, we construct an $m\times m$ matrix $\mathbf{R}_{Bayes}$ with columns as small areas, rows as ranks, and the $(i, j)$-th entry containing the  relative frequency  the $i$-th small area has rank $j$.  
  We obtain $\mathbf{R}_{Bayes}$ as follows:  starting with an empty matrix,  for any  $\boldsymbol{\theta}^{(s)}$ in $\mathcal{S}_{Bayes, \alpha}$, if the $j$-th entry (i.e., small area) has the $i$-th rank, we  add one in the $(i, j)$-th slot, for $j = 1, 2, \ldots, m$. Thus we get a matrix where every row and every column has exactly $(m-1)$ entries as zero and exactly one entry as one, consequently the matrix is \textit{a fortiori} doubly stochastic. The process is repeated for all elements of $\mathcal{S}_{Bayes, \alpha}$, and  $\mathbf{R}_{Bayes}$ is the average of these matrices,  so it remains doubly stochastic. 
  The $j$-th column of  $\mathbf{R}_{Bayes}$, denoted by  $\mathbf{R}_{Bayes, j}$, corresponds to the relative frequencies of the ranks $\{1, 2, \ldots, m \}$ that the $j$-th small area may take, when $\boldsymbol{\theta} \in \mathcal{S}_{Bayes, \alpha}$. 
Table~\ref{tab:HB.HPD.9States_251021} is an example of $\mathbf{R}_{Bayes}$ for the 9 states in Example~\ref{Exmp:1} that have the lowest commuting times. 
  Such a doubly stochastic matrix  is not available for KWW method, a limitation of their approach. 
However, suppose the $j$-th small area is assigned $m_{j} \leq m$ possible ranks in $\mathcal{R}_{KWW, \alpha}$. We may construct an $m$-dimensional vector $\mathbf{R}_{KWW, j}$ containing $1/m_{j}$ in the entries corresponding to the ranks it can take and zero otherwise, and define the matrix $\mathbf{R}_{KWW}$ by stacking these as columns. For easier reproducibility, we have also presented our proposed approach as a step-by-step procedure in supplementary Section~\ref{Sec:Small_m}.

\subsection{Interpretation of ranks based on credible sets}
\label{Sec:Interpretation}

Since there are several steps in the process of computing $\mathbf{R}_{Bayes}$ and the relative frequencies  $\mathbf{R}_{Bayes, j}$'s of the $j$-th small area, and there are multiple sources of randomness in our Bayesian framework, a natural question is how to interpret the results. This is also relevant for the numerous applications of small area rankings at NSOs and elsewhere. We consider $\alpha \in (0, 1)$ here, the case of $\alpha = 0$ is treated separately in Section~\ref{Sec:DirectRanking}.

A simple way of interpreting $\mathbf{R}_{Bayes, j}$  is in terms of bounds for probabilities, moments and support of the posterior ranks of the $j$-th small area. Consider for example, the $\mathbf{R}_{Bayes}$ matrix presented in Table~\ref{tab:HB.HPD.9States_251021} on the 9  states from Example~\ref{Exmp:1} that have the lowest commuting times ($\mathbf{Y}$), with $\alpha =0.1$. We see that $\mathbf{R}_{Bayes, ID}$ (relative weights of ranks of ID, when $\boldsymbol{\theta}$ is restricted to $ \mathcal{S}_{HB, E, \alpha}$) is supported on the ranks 8 and 9 with values 0.0003 and 0.9997 respectively, and the interpretation is that  \textit{with a posterior probability of at least 90\% ($= 1- \alpha$), the ranks that ID can take are 8 and 9, and the posterior probability that the rank of ID is 9 is at least 0.89973.} To see these, partition the parameter space in terms of the 
credible set and its complement, that is $\boldsymbol{\Theta}_{m} = \mathcal{S}_{Bayes, \alpha} \cup \mathcal{S}_{Bayes, \alpha}^{C}$, where $A^{C}$ denotes the complement of a set $A$. Now notice that $P[ \textrm{Rank} (ID) = 9] \geq P[ \textrm{Rank} (ID) = 9 | \mathcal{S}_{Bayes, \alpha}] P [\mathcal{S}_{Bayes, \alpha}] \geq 0.9997 (1 - \alpha) = 0.89973$. Similarly, we can claim that the posterior probability that the rank of SD is 1, 2, or 3 is at least 90\%, and the Bayes estimator (posterior mean) of its rank is within (1.38, 2.18), or that the posterior probability is at least 90\% that the lowest commuting times are in SD, ND or WY, but the chance that WY has the lowest commuting time is at most 0.100027. Such inequalities naturally are sharpened as $\alpha \downarrow 0$. Additional interpretations will be pursued in future research.

It should also be noted that the ranks of small areas are typically used by NSOs and other practitioners in conjunction with the predicted values of $\boldsymbol{\theta}$; the latter containing pertinent information about the quantities of importance to policymakers and stakeholders. Also, the ranks $\check{\mathbf{r}}$ are functions of $\boldsymbol{\theta}$, and the set of all possible ranks is discrete and has super-exponential growth. Consequently uncertainty quantification of $\check{\mathbf{r}}$ directly, without referencing the uncertainties in the prediction of $\boldsymbol{\theta}$, is both computationally inefficient and redundant, since uncertainties in $\hat{\boldsymbol{\theta}}$ have to be taken into account anyway by NSOs in their downstream tasks. In view of this issue, KWW constructed a confidence set for $\boldsymbol{\theta}$ first to obtain the confidence set for $\check{\mathbf{r}}$; similarly, we obtain credible sets of $\boldsymbol{\theta}$ and utilize those for uncertainty quantification of $\check{\mathbf{r}}$. 

Also note that the use of relative frequencies within a data-driven set for statistical inferences is not new, although our usage of it in a Bayesian context may be new. For example, in a model selection context, \cite{liu2021simple} in their bootstrap Monte Carlo construction of an approximate $(1-\alpha)$ confidence set for the {\it true model number} (a discrete parameter of interest, similar to our paper), sequentially add models to the confidence set until the empirical coverage of the set of selected models is at least $(1-\alpha)$. Each of the elements (models) included in the confidence set has an empirical probability of being the {\it true model;} see equations (3)-(5), page 677, of \cite{liu2021simple}. See also \cite{li2024measures} for another related model selection approach where the underlying principle is similar.

\subsection{The $\alpha = 0$ case: Ranking directly from the MCMC samples}
\label{Sec:DirectRanking}

Readers may wonder why we are constructing a credible set before ranking the $\boldsymbol{\theta}$ vectors, which is undoubtedly an extra computational step. First, skipping the credible set construction step (thus effectively setting $\alpha = 0$), results in inclusion of all MC (or MCMC) samples, some of which may be extreme outliers and present in very low posterior probability regions. This results in an obvious practical robustness problem, since the rank vector from such a $\boldsymbol{\theta}$ may not reflect its extreme nature at all. Second, it is overwhelmingly likely that rank vectors based on  MCMC samples are all unique, and provide no useful summarization at all. To appreciate this, consider the commute data  Example~\ref{Exmp:1}. The number of possible rank vectors for the 51 small areas  is $51! = O (10^{66})$, about 2.5 times the size of the observed universe in meters. Even an extremely high choice of MCMC sample size, say $10^{10}$, would result in each realized rank vector to have vanishingly small probability $O (10^{-56})$. Consequently, meaningful inference using the empirical rank vectors from the MCMC samples involving all the small areas is impossible. Third, in upending, the importance of $\boldsymbol{\theta}$ vectors and their ranks, and treat the latter as the primary objects of interest, we ignore the reality that policy guidance and stakeholder interests are primarily served with prediction, inference and uncertainty quantification of $\boldsymbol{\theta}$ conditional on the data. The ranks of the small areas and associated relative weights  are of importance in ensuring fairness, in making sure that chance variations do not significantly impact fund allocations, downstream projects and other tasks, and hence we study this in our paper, but a focus on the ranks should not distract us from the fact that these are coupled with actual $\boldsymbol{\theta}$ values that inform the actual policies. Hence, this section is for conceptual clarity.

\begin{table}[h]\caption{Matrix $\mathbf{R}_{HB}\times 100$ for the 9 states from Example~\ref{Exmp:1} that have the lowest commuting times ($\mathbf{Y}$), based on empirical posterior distribution of rank vector based on entire MCMC sample in the HB framework.}\label{tab:HB.9States}
\begin{center}
{\scriptsize
\begin{tabular}{|c|rrrrrrrrr|}
      \hline
     & \multicolumn{9}{c|}{States} \\
Rank & ID & KS & IA & AK & MT & NE & WY & ND & SD \\
\hline
1&      &      &      &      &      &      & 0.38 & 45.08 & 54.54 \\
2&      &      &      & 0.09 & 0.31 & 0.17 & 2.53 & 52.48 & 44.42 \\
3&      &      &      & 8.33 & 22.06 & 31.17 & 35.14 & 2.27 & 1.02 \\
4&      & 0.02 & 0.03 & 14.93 & 26.97 & 39.04 & 18.84 & 0.13 & 0.02 \\
5&      & 0.98 & 1.10 & 27.70 & 27.93 & 22.75 & 19.51 & 0.03 &      \\
6&      & 6.65 & 12.20 & 38.38 & 19.45 & 6.87 & 16.44 &      &      \\
7& 0.04 & 30.82 & 56.34 & 6.95 & 2.24 &      & 3.60 &      &      \\
8& 0.51 & 61.02 & 30.31 & 3.61 & 1.02 &      & 3.53 &      &      \\
9&99.44 & 0.51 & 0.02 &      &      &      & 0.02 &      &      \\
\hline
\end{tabular}
}
\end{center}
\end{table}

Also note that if the number of small areas under study is not very large, then  a full-scale computation for inference of ranks in the $\alpha = 0$ setting is viable, but does not produce any different insights from computations where a small positive value of $\alpha$ is used. For example, corresponding to Table~\ref{tab:HB.HPD.9States_251021} above with $\alpha = 0.1$, in Table~\ref{tab:HB.9States} we present the $\alpha = 0$ case. 
 The differences in the entries of these two tables are minor; and using only the $\boldsymbol{\theta}$ samples from a credible set does not significantly alter the downstream inferences about posterior probabilities and expectations of the ranks of various small areas. 
Our proposed approach, where we first construct a $(1 - \alpha)$-credible set for $\boldsymbol{\theta}$ irrespective of its dimension (number of small areas) and then rank the entries within this credible set, allows for \textit{simultaneous inference} relating to $\boldsymbol{\theta}$'s and their ranks.  Each rank vector that appears in subsequent calculations can be guaranteed to be associated with a predicted $\boldsymbol{\theta}$ value that has a certain level of credibility. Naturally, some care is needed in interpreting the results from such simultaneous inference, and we discuss this in greater detail in Section~\ref{Sec:Interpretation}.

\section{A simulation study using a baseball example }\label{Sec:Simulation}

For this section and the next two sections, we fix $\alpha = 0.1$, and the results should be interpreted along the lines stated in Section~\ref{Sec:Interpretation}; however, they can be equivalently recomputed 
and interpreted with $\alpha = 0$ also. We skip the details for the $\alpha =0$ case, since the results and broad conclusions are similar to the $\alpha = 0.1$ results presented below.

\begin{table}[h]\caption{Results from simulations based on \cite{EfronMorris1975} framework. For each value of $A$, the  rows respectively contain the average of posterior mean absolute deviations $\overline{AMD}$, the $m^{-1}$-th power of volume ($V^{1/m}$) and length ($L$) of the credible/confidence set of $\boldsymbol{\theta}$ and average mass concentration ($\overline{EC}$).
}\label{baseball-sim}
		\begin{center}
			{\scriptsize
				\begin{tabular}{|c|c|rrrrrr|rrrrrr|}
					\hline
& & \multicolumn{6}{c|}{With covariate} & \multicolumn{6}{c|}{Without covariate} \\
& & & & \multicolumn{2}{c}{UB} & \multicolumn{2}{c|}{HB} &  & & \multicolumn{2}{c}{UB} & \multicolumn{2}{c|}{HB} \\
$A$    &              &   KWW & MRSW &   C  &  E &   C &   E  &   KWW & MRSW &  C  &  E &   C &   E  \\
\hline
\multirow{4}*{0.001} & $\overline{AMD}$ &   5.49 & 4.99 & 2.51 & 2.26 & 1.52 & 1.17 & 5.98 & 5.95 & 4.81 & 4.65 & 5.18 & 5.33 \\
                     & $V^{1/m}$        &   0.36 &      & 0.36 & 0.29 & 0.27 & 0.21 & 0.36 &      & 0.36 & 0.29 & 0.26 & 0.21 \\
                     & $L$                &   0.36 &      & 0.36 & 0.29 & 0.25 & 0.20 & 0.36 &      & 0.36 & 0.29 & 0.24 & 0.20 \\
                     & $\overline{EC}$  &   16.93 & 15.71 & 8.50 & 6.60 & 5.71 & 3.18 & 17.99 & 17.92 & 12.75 & 10.69 & 15.67 & 16.25 \\
\hline
\multirow{4}*{0.005} & $\overline{AMD}$ &   5.24 & 4.71 & 2.38 & 2.15 & 2.00 & 1.80 & 5.91 & 5.78 & 3.75 & 3.50 & 4.13 & 4.11 \\
                     & $V^{1/m}$        &   0.36 &      & 0.36 & 0.29 & 0.30 & 0.24 & 0.36 &      & 0.36 & 0.29 & 0.29 & 0.24 \\
                     & $L$                &   0.36 &      & 0.36 & 0.29 & 0.29 & 0.23 & 0.36 &      & 0.36 & 0.29 & 0.28 & 0.23 \\
                     & $\overline{EC}$  &   16.35 & 15.04 & 7.96 & 6.15 & 6.97 & 4.95 & 17.85 & 17.54 & 11.27 & 9.18 & 13.43 & 12.91 \\
\hline
\multirow{4}*{0.010} & $\overline{AMD}$ &   4.99 & 4.43 & 2.25 & 2.04 & 2.09 & 1.88 & 5.71 & 5.39 & 3.01 & 2.75 & 3.24 & 3.01 \\
                     & $V^{1/m}$        &   0.36 &      & 0.36 & 0.29 & 0.32 & 0.25 & 0.36 &      & 0.36 & 0.29 & 0.31 & 0.25 \\
                     & $L$                &   0.36 &      & 0.36 & 0.29 & 0.31 & 0.25 & 0.36 &      & 0.36 & 0.29 & 0.31 & 0.25 \\
                     & $\overline{EC}$  &   15.74 & 14.28 & 7.57 & 5.86 & 7.29 & 5.55 & 17.39 & 16.64 & 9.73 & 7.73 & 11.07 & 9.50 \\
\hline
\multirow{4}*{0.100} & $\overline{AMD}$ &   2.97 & 2.53 & 1.22 & 1.09 & 1.22 & 1.09 & 3.07 & 2.60 & 1.29 & 1.16 & 1.30 & 1.16 \\
                     & $V^{1/m}$        &   0.36 &      & 0.36 & 0.29 & 0.35 & 0.28 & 0.36 &      & 0.36 & 0.29 & 0.35 & 0.28 \\
                     & $L$                &   0.36 &      & 0.36 & 0.29 & 0.35 & 0.28 & 0.36 &      & 0.36 & 0.29 & 0.35 & 0.28 \\
                     & $\overline{EC}$  &   10.22 & 8.75 & 4.43 & 3.46 & 4.49 & 3.50 & 10.49 & 8.94 & 4.60 & 3.60 & 4.68 & 3.66 \\
\hline
\multirow{4}*{1.000} & $\overline{AMD}$ &   1.23 & 1.02 & 0.46 & 0.40 & 0.46 & 0.40 & 1.23 & 1.04 & 0.45 & 0.39 & 0.45 & 0.39 \\
                     & $V^{1/m}$        &   0.36 &      & 0.36 & 0.29 & 0.36 & 0.29 & 0.36 &      & 0.36 & 0.29 & 0.36 & 0.29 \\
                     & $L$                &   0.36 &      & 0.36 & 0.29 & 0.36 & 0.29 & 0.36 &      & 0.36 & 0.29 & 0.36 & 0.29 \\
                     & $\overline{EC}$  &   4.43 & 3.75 & 2.11 & 1.79 & 2.12 & 1.80 & 4.47 & 3.79 & 2.11 & 1.78 & 2.11 & 1.78 \\
\hline
				\end{tabular}
			}
		\end{center}
	\end{table}

Table \ref{baseball-sim} below uses various summary measures to compare Bayesian solutions with the KWW frequentist solution. For these measures that are extensively described in \ref{Sec:Compare}, we start with a doubly stochastic matrix $\mathbf{R}_{Bayes}$, the entries of its $j$-th column $\{ \pi_{i, j, \mathbf{Y}} = \mathbf{R}_{Bayes, i, j}: i = 1, \ldots, m \} $ denote the relative weights on the ranks for the $j$-th small area. From this an estimated rank of the $j$-th small area is the mean $\hat{{r}}_{j} = \sum_{i =1}^{m} i \pi_{i, j, \mathbf{Y}}$. We quantify the uncertainty of $\mathbf{R}_{Bayes, j}$ in capturing the true or estimated rank using the posterior \textit{mean absolute deviation} defined as 
$AMD_{j} = \sum_{i =1}^{m} | i - \check{{r}}_{j}| \pi_{i, j, \mathbf{Y}}$ if $\check{{r}}_{j}$ is known (as in simulations or in the presence of ground truth), and $\widehat{AMD}_{j} = \sum_{i =1}^{m} | i - \hat{{r}}_{j}| \pi_{i, j, \mathbf{Y}}$ otherwise. Overall uncertainties are measured using $\overline{AMD} = m^{-1} \sum_{j = 1}^{m} AMD_{j}$ and 
$\widetilde{AMD} = m^{-1} \sum_{j = 1}^{m} \widehat{AMD}_{j}$. 
To quantify the spread of weights associated with ranks as in $\mathbf{R}_{Bayes, j}$, we use \textit{effective cardinality}, defined as $EC_{j} = \exp[ - \sum_{i =1}^{m} \pi_{i, j, \mathbf{Y}} \log (\pi_{i, j, \mathbf{Y}})]$, the exponential of the entropy. Overall spread is quantified with $\overline{EC} = m^{-1} \sum_{j=1}^{m} EC_{j}$. Smaller values of $\overline{EC}$ imply the distribution is more concentrated, hence more informative for predictions and inferences. If $\pi_{i, j, \mathbf{Y}} \equiv m^{-1}$ we have $EC_{j} = m$, the cardinality of the set of ranks.

\cite{EfronMorris1975} considered batting averages of 18 major league baseball batters from their first 45 at-bats in 1970 to predict their performances in the rest of the season. We emulate this framework  
in simulations to compare KWW and MRSW with our four Bayesian proposals. We generate 18 $x_i$'s independently from a uniform (0,1) distribution to use as covariates for the simulation, we keep these fixed throughout. Then, we generate  independent $\theta_i$'s from $N(\beta_0 +\beta_1 x_i, A)$ 
and $Y_i$ from $N(\theta_i, D_i)$. We use the $D_i$'s from \cite{EfronMorris1975}, $A = 0.001, 0.005, 0.01, 0.1$ and $1$ (for the baseball data $\hat A \approx 0.002$),
and  $\beta_1 =0$ (no covariate) and $\beta_1 = 0.4$. 
We fix $\beta_0 =0.2$, but $\beta_0$ has no influence on ranks.
 Table~\ref{baseball-sim} contains the average of the posterior mean absolute deviations of the components of the overall ranks  {($\overline{AMD}$)}, {the $1/m$-th root of} average of the volumes ($V^{1/m}$) of credible/confidence sets,  average length ($L$) and effective  cardinality $\overline{EC}$. 
 In terms of all the measures, the HB elliptical method is typically the best performer in the presence of covariates. All four Bayesian methods significantly outperform the frequentist approaches KWW and MRSW in terms of concentration measures {($\overline{AMD}$)} and $\overline{EC}$, both in presence or absence of covariates. 
 When the $\theta_i$'s vary according to a covariate, the HB method performs substantially better for the smaller values of $A$ (0.001, 0.005, 0.01) relative to the $D_i$'s, as in these cases there is a substantial structure in the $\theta_i$'s.
 In all situations, the four Bayesian techniques overwhelmingly outperform KWW and MRSW, typically by a factor of two to three when a strong with covariate model holds. 


\section{Ranking of the US states based on commuting times}\label{kww-commuting}

\begin{figure}[htb]\caption{The  90\% joint confidence region for commute time ranking (replicating Figure 1 of KWW), with  $\mathcal{R}_{UB, C, \alpha}$-based Bayesian ranks with associated weights overlay. The corresponding figure using the MRSW approach instead is contained in Figure~\ref{Fig:Commute_SP} of the supplementary materials.}\label{Fig:Commute}
\begin{center}
    \includegraphics[width=.49\textwidth,angle=90]{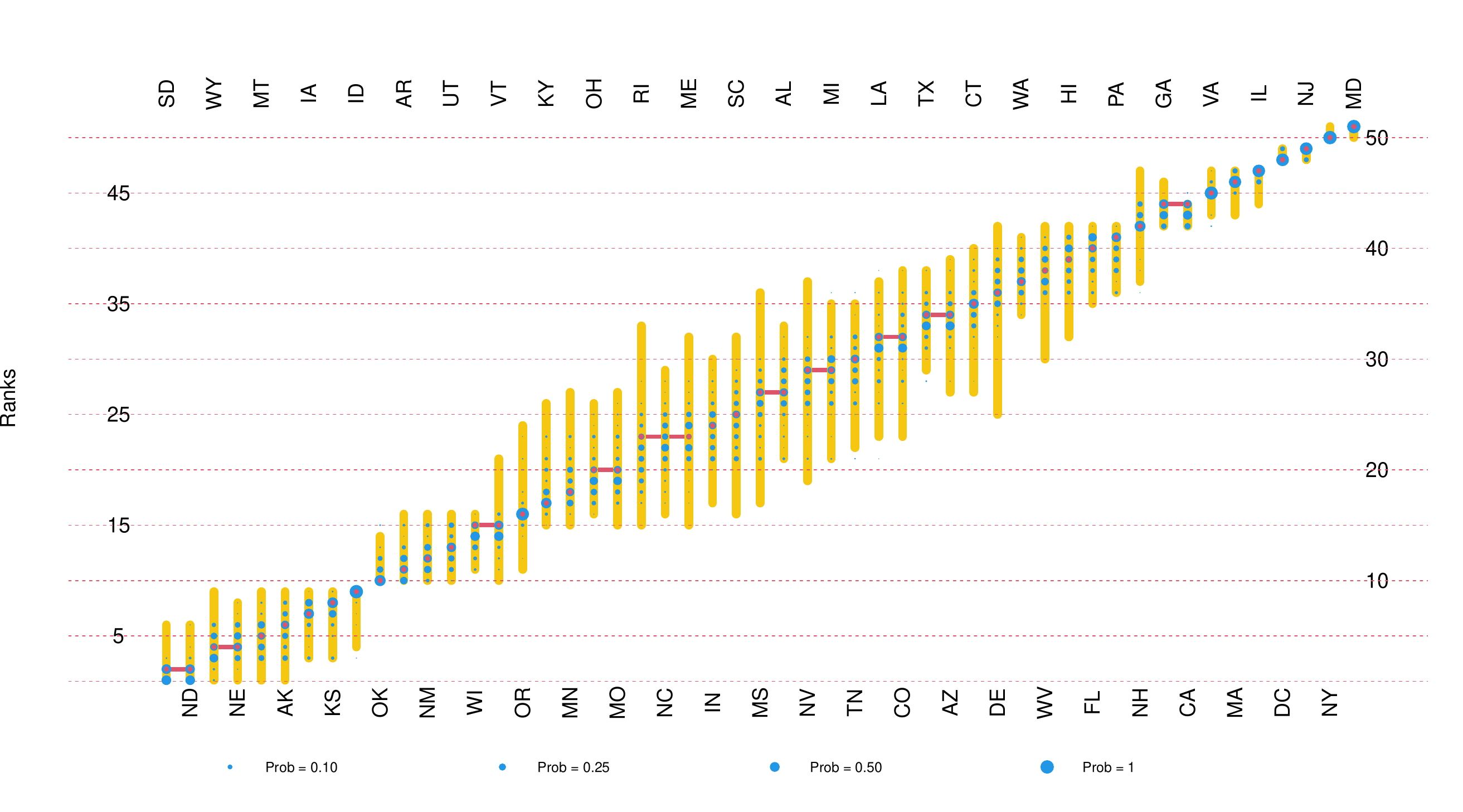} 
\end{center}
\end{figure}

The pioneering  \cite{KWW2020} article  used frequentist approach for the problem presented in Example~\ref{Exmp:1}, to rank the fifty states of the U.S. and DC by mean commuting times of workers sixteen or older and not working from home. They used survey data collected from the ACS. In Figure~\ref{Fig:Commute}, we recreate Figure~1 of \cite{KWW2020} that depicts $\mathcal{R}_{KWW, \alpha}$. In this example, we fix $\alpha = 0.1$, for comparison with KWW. Possible marginal ranks for state $j$ are shown by a yellow line segment that extends from $|\Lambda_{Lj}| +1$ to $|\Lambda_{Lj}| +1+|\Lambda_{Oj}|$ {(= $m-|\Lambda_{Rj}|$)}. For example, the possible ranks for the state ID are $4$ to $9$. On the other hand, the states that can have the rank 9 are WY, AK, MT, IA, KS and ID (the line segments for these states intersect the horizontal line for rank 9). We use a red dot at $r_{j} \in \{1, \ldots, m \}$ to depict that $y_{j}$ has rank $r_{j}$ in the observed data $\mathbf{y}$. If two states tie for ranks $r, r+1$, their tied ranks are shown by a red line segment joining these states. For example, SD and ND  tie for ranks 1 and 2.

The Bayesian analog of the KWW approach is $\mathcal{R}_{UB, C, \alpha}$. To illustrate the efficacy of this  minor Bayesian modification, 
we overlaid on Figure~\ref{Fig:Commute} with $\mathbf{R}_{UB, C, j}$-based ranks for each state using blue circles with sizes proportional to the weights. To better understand this figure, let us concentrate on NE. From $\mathcal{R}_{KWW, \alpha}$, we observe that the rank set for NE is 1-8, and moreover, each of these ranks is equally viable. In comparison, from $\mathbf{R}_{UB, C, NE}$, we observe that NE has ranks $3-6$ with relative frequencies $0.21$, $0.20$, $0.55$ and $0.04$, respectively.  Additionally, Figure~1 of \cite{KWW2020} states SD, ND, WY, NE, MT, AK, IA, KS and ID (the first nine states depicted along the $x$-axis of Figure~\ref{Fig:Commute}) all can have rank 4 with equal certainty, while  we obtain that the states that can plausibly have rank 4 are NE, MT, WY and AK (with respective relative frequencies $0.55$, $0.23$,  $0.18$ and $0.04$). Thus  $\mathbf{R}_{UB, C, j}$ shows that NE has ranks $3-6$ with a relative weight of \textit{at least} $90\%$. Further, the posterior mean of its rank is bounded between 4.078 and 9.078.  
Supplementary Figure~\ref{Fig:Commute_SP} has the 90\% confidence region using the MRSW approach. As is evident, there is little difference between the outputs of KWW and MRSW, with both these non-Bayesian techniques typically resulting in wider bands of ranks for most US states. The bands from MRSW are sometimes marginally  narrower than those from KWW, but nevertheless are non-competitive with Bayesian approaches. This can also be seen from Table~\ref{Tab:Supple_NumerOfRankVectors}, where we considered the $k$  states with lowest commute times sequentially, for $k = 4, \ldots, 10$ and counted the number of rank vectors that are included in the 90\% confidence/credible sets  
$\mathcal{R}_{KWW, 0.1}$, $\mathcal{R}_{MRSW, 0.1}$, $\mathcal{R}_{UB, C, 0.1}$, $\mathcal{R}_{UB, E, 0.1}$, $\mathcal{R}_{HB, C, 0.1}$, $\mathcal{R}_{HB, E, 0.1}$. Notice that the both $\mathcal{R}_{KWW, 0.1}$ and $\mathcal{R}_{MRSW, 0.1}$ rapidly grow in size compared to the credible sets, the former growing extremely fast. This suggests that even for applications involving moderate number of small areas, using the non-Bayesian approaches may not be computationally viable.


The KWW-based ranks $\mathbf{R}_{KWW, j}$'s can also lead to impractical and implausible solutions. To see this, notice  from Table~\ref{tab:HB.HPD.9States_251021} that it is practically certain that ID ranks 9, AK ranks $3-8$, and SD ranks 1 or 2. The corresponding KWW sets, respectively, are $\{4-9\}, \{1-9\}$ and $\{1-6\}$.  
Thus from the $\mathbf{R}_{KWW, j}$'s, the rank assignment 4 for ID, 5 for AK and 6 for SD is as probable as any other choice. However, this assignment is strongly contradicted by the double scatter plot in Figure~\ref{FigS1:ID_AK_SD}, which makes it abundantly clear that the commute time for ID is more than for AK, which is more than for SD.

\begin{table}[h]\caption{Commute time example: Different comparative measures on the 90\% confidence sets $\mathcal{S}_{KWW, 0.1}$, $\mathcal{S}_{MRSW, 0.1}$  and the 90\% partially ordered credible sets $\mathcal{S}_{UB, C, 0.1}$, $\mathcal{S}_{UB, E, 0.1}$, $\mathcal{S}_{HB, C, 0.1}$, and $\mathcal{S}_{HB, E, 0.1}$. }\label{comp_sets_commu-text_251021}
	\begin{center}
		{\scriptsize
			\begin{tabular}{|cc|c|c|cc|cc|}
				\hline
				&                              &&         & \multicolumn{2}{c|}{UB} & \multicolumn{2}{c|}{HB} \\
				Parameter & Measure                      &   KWW    & MRSW            &  Cartesian           & Elliptical           & Cartesian            & Elliptical \\
				\hline
				$\boldsymbol{\theta}$  & Volume          & 4.52     &           & 2.57                 & $1.71\times 10^{-7}$ & 3.85                 & $1.53\times 10^{-7}$ \\
				$\boldsymbol{\theta}$  & Average length  & 1.16     &           & 1.15                 & 0.83                 & 1.16                 & 0.83  \\
				\hline
                $\check{\mathbf{r}}$& $\widetilde{AMD}$  & 2.49     & 2.21      & 0.78                 & 0.60                 & 0.78                 & 0.62 \\
				 $\check{\mathbf{r}}$& $\overline{EC}$      &10.02     & 8.92      & 3.89                 & 2.97                 & 3.90                 & 3.03 \\
				\hline
			\end{tabular}
		}
	\end{center}
\end{table}

We repeated the analysis using the HB framework as well as elliptical credible sets.
The ranking results using $\mathbf{R}_{HB, C, j}$'s, $\mathbf{R}_{HB, E, j}$'s or $\mathbf{R}_{UB, E, j}$ are very similar to those using $\mathbf{R}_{UB, C, j}$'s for all small areas, showing that our Bayesian approach is quite robust against prior choices and the shape of the credible sets. However, there is a difference across the sizes of the {credible sets}, and hence the uncertainty associated with the Bayesian rankings. We report the details in Table~\ref{comp_sets_commu-text_251021}. From this table, we find that no credible set has larger volume than $\mathcal{S}_{KWW, \alpha}$. Note that MRSW provides neither point estimates nor confidence sets for $\boldsymbol{\theta}$. This is a major limitation of their approach, as prediction, inference, and uncertainty quantification for $\boldsymbol{\theta}$ are the primary drivers of small area statistics.

Although the volumes of Cartesian credible sets are of similar order as the volume of $\mathcal{S}_{KWW, \alpha}$, the elliptical credible sets are approximately $10^{-7}$ order of magnitude smaller.  The $\widetilde{AMD}$ and $\overline{EC}$ for KWW and MRSW are {at least twice} of the Bayesian counterparts, with elliptical credible sets being distinctly better.  Thus, our strong recommendation to practitioners is to use the HB framework and elliptical credible sets.
Our framework can also be used to identify entities or regions that are among the top (or bottom) $x\%$ by computing appropriate posterior probabilities. For example, we can obtain the most likely subsets of five (10\%) states from the 51 states with the longest mean commute times. The subsets (MD, NY, NJ, DC, IL), (MD, NY, NJ, DC, MA) and (MD, NY, NJ, DC, VA) have longest five mean commute times  with respective weights 88.11, 11.84 and 0.04, see supplementary Table~\ref{Tab:Top_Ranks}. {Similarly, for the six Central Atlantic states MD, NY, NJ, DC, VA, and PA, only nine out of  ${51 \choose 6} \times 6!$ rank-vectors together account for more than 97\% of the total weight, when using a 90\% elliptical credible set in the hierarchical Bayesian framework, see 
supplementary Table~\ref{Tab:Rank_dist} and its graphical display in Figure~\ref{Fig:RankComb} for details.} 

\section{Ranking of the US states based on median incomes}\label{medianincome}

\begin{figure}[h]\caption{The  90\% joint confidence region for income ranking using KWW, with  $\mathcal{R}_{HB, E, 0.1}$-based Bayesian ranks with associated weights overlay. The corresponding figure  using MRSW approach, with the same overlay is presented in Figure~\ref{Fig:Income_SP} of the supplementary materials.}\label{Fig:Income}
\begin{center}
    \includegraphics[width=.97\textwidth]{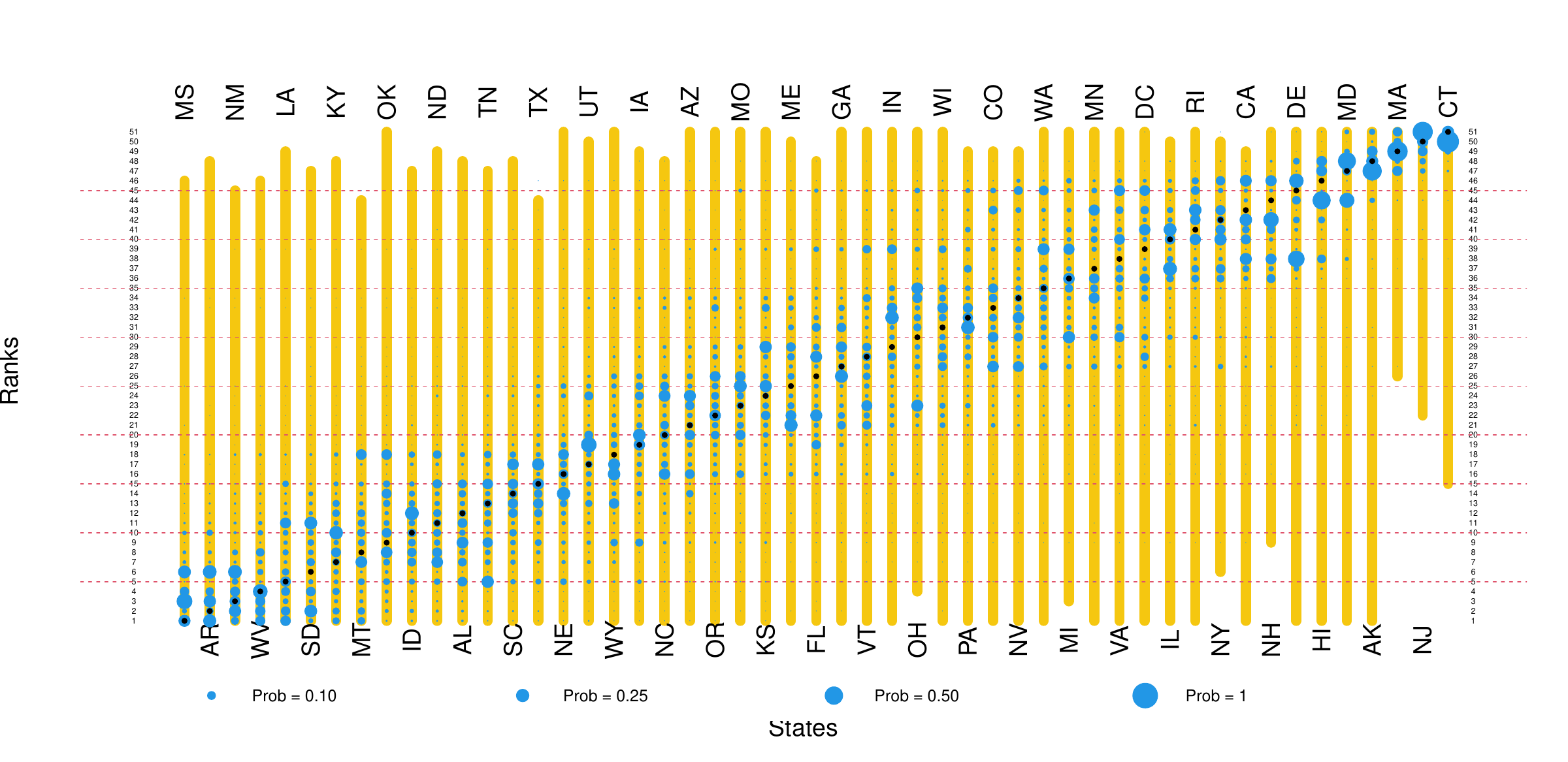} 
\end{center}
\end{figure}

The 1989 median incomes for the US states reported in the 1990 census are considered accurate. Treating these as ``gold standard", this problem is considered a benchmark for evaluating effectiveness of small area models/methods; see for example, \cite{Ghosh1996EstimationOM} and \cite{ChungDatta2022}.
  The Current Population Survey  provides annual estimates of median incomes for states for various family sizes, which we use as the response variable, with covariates  $x_{i1} =$ the $i$th state median income for 1979 from the 1980 Census, and 
$x_{i2} =(\text{PCI}_{i,1989}/\text{PCI}_{i,1979})x_{i1}$, $i=1,\ldots,m$, where $\text{PCI}_{i,1979}$ and $\text{PCI}_{i,1989}$ are the 1979 and 1989 per capita incomes (PCI) of the $i$th state provided by the Bureau of Economic Analysis of the U.S. Department of Commerce. 
We analyze this data using the two frequentist algorithms KWW and MRSW, and our four Bayesian proposals, as in Section~\ref{kww-commuting}, but this time we also have the ``gold standard'' for evaluating  relative performances.

\begin{table}[h]\caption{Median income data {in terms of 10,000\$}: Different comparative measures on the 90\% confidence sets $\mathcal{S}_{KWW, 0.1}$, $\mathcal{S}_{MRSW, 0.1}$   and the 90\% partially ordered credible sets $\mathcal{S}_{UB, C, 0.1}$, $\mathcal{S}_{UB, E, 0.1}$, $\mathcal{S}_{HB, C, 0.1}$, and $\mathcal{S}_{HB, E, 0.1}$.}\label{medianincomes_251022}

	\begin{center}
		{\scriptsize
			\begin{tabular}{|cc|c|c|cc|cc|}
				\hline
				&                              &&         & \multicolumn{2}{c|}{UB} & \multicolumn{2}{c|}{HB} \\
				Parameter & Measure                      &   KWW    & MRSW            &  Cartesian           & Elliptical           & Cartesian            & Elliptical \\
				\hline
				$\boldsymbol{\theta}$  & Volume          & $1.08\times 10^{27}$ &       &  $6.15\times 10^{26}$ & $4.08\times 10^{19}$ & $7.05\times 10^{19}$ & $3.65\times 10^{12}$ \\
				$\boldsymbol{\theta}$  & Average length  &  3.59                &       &  3.55                 & 2.57                 & 2.50                 & 1.79  \\
                \hline
				$\check{\mathbf{r}}$& $\widetilde{AMD}$  & 11.98                & 11.51 &  4.92                 & 3.88                 & 4.00                 & 2.42 \\
				$\check{\mathbf{r}}$& $\overline{AMD}$   & 15.67                & 14.91 &  8.02                 & 7.60                 & 5.10                 & 3.39 \\
				 $\check{\mathbf{r}}$ & $\overline{EC}$  & 47.94                & 46.06 & 23.10                 &15.13                 &19.53                 & 9.85 \\
				\hline
			\end{tabular}
		}
	\end{center}
\end{table}

 Figure~\ref{Fig:Income} (supplementary Figure~\ref{Fig:Income_SP}) is analogous to Figure~\ref{Fig:Commute} (Figure~\ref{Fig:Commute_SP}), however, this time the blue circles are based on our recommended algorithm $\mathcal{R}_{HB, E, \alpha}$; recall that we fix $\alpha = 0.1$ throughout this section. It can be seen that both KWW and MRSW have wide bands of ranks for most states. On the other hand, $\mathcal{R}_{HB, E, \alpha}$ is compact and clearly more informative and useful to policymakers, planners, and stakeholders. Table~\ref{medianincomes_251022} reports various properties of the credible sets and relative frequencies in this example, and again it is evident that all Bayesian algorithms outperform the non-Bayesian ones, with HB and {elliptical credible set} being the best of the lot. Additional results for this example are presented in supplementary Section~\ref{Sec:Supple_Tables}, where we also show that the performance of MRSW is often better than KWW, but nowhere competitive with $\mathcal{R}_{HB, E, \alpha}$, and generally not competitive with the Bayesian approaches.

\section{The $\Delta$-FH framework and some theoretical studies}
\label{Sec:Theory}

Notice that the posterior distribution of $\boldsymbol{\theta}$ from either Section~\ref{unstrBayes} or Section~\ref{HBayes} does not concentrate anywhere; that is, there does not exist a location in $\boldsymbol{\Theta}_{m}$ and any increasing sequence (in $m$) of scale  such that the either of these posterior concentrates. Similar lack of convergence or consistency can be easily observed for frequentist predictors of  $\boldsymbol{\theta}$. For example, the empirical best linear unbiased predictor (EBLUP) $\hat{\boldsymbol{\theta}}$ satisfies $\hat{\boldsymbol{\theta}} - {\boldsymbol{\theta}} = O_{P} (1)$ for all $m$ except for pathological cases that have no bearing on any realistic small area problems. In view of these issues, we propose a triangular-Fay-Herriot ($\Delta$-FH in notations) model, and establish interesting properties of this model that demonstrate the performance of the Bayesian ranking technique presented in this paper.

In the following, we consider a triangular array of data that is modeled using a Fay-Herriot model at every stage. At the $n$-th stage, the data is given by $\mathbf{Y}_{n} = (Y_{n, 1}, \ldots, Y_{n, m_{n}})^{T}$, and $Y_{n, i}$ is associated with covariates $\mathbf{x}_{n, i} \in \mathbb{R}^{p}$ and a known sampling variance $D_{n, i}$. The unobserved latent quantity of interest is $m_{n}^{-1/2} \boldsymbol{\theta}_{m_{n}} \in \boldsymbol{\Theta}_{m_{n}}$, where $\{ m_{n} \}$ is a strictly increasing sequence of positive integers denoting the dimension of $\boldsymbol{\Theta}_{m_{n}}$. We require the factor of $m_{n}^{-1/2}$ associated with $\boldsymbol{\theta}_{m_{n}}$  to ensure it is a $O_{P} (1)$ random variable in the theoretical analysis. For clarity in exposition and to avoid complicated subscripts below, we henceforth assume without loss of generality that $m_{n} \equiv n$. Also in the following, we will drop the subscript ``${n}$'' when there is no scope for confusion, and consequently often denote the data as $\mathbf{Y}_{n} = (Y_{1}, \ldots, Y_{n})^{T}$, the $i$-th covariate as $\mathbf{x}_{i}$, the $i$-th sampling variance as $D_{i}$ and so on. 

The $\Delta$-Fay-Herriot model is given by the
($i$) \textbf{sampling model:} \ \  $Y_{n, i} | \theta_{n, i}  \stackrel{ind}{\sim}  N(\theta_{n, i}, D_{n, i})$, and the $(ii)$
\textbf{linking model: } $ \theta_{n, i}| \boldsymbol{\beta}_{n}, A_{n} \stackrel{ind}{\sim}  N(\mathbf{x}_{n, i}^{T}\boldsymbol{\beta}_{n}, A_{n})$
independently for $i = 1, \ldots, {n}$.  The parameters $\{ \boldsymbol{\beta}_{n} \in \mathbb{R}^{p}: n \geq 1 \}$ and $\{ A_{n} > 0: n \geq 1 \}$ are unknown and have prior distributions. The case where these parameters are fixed and constant for all $n$ is a special case. Further, we use the notation $\mathbf{D}_{n} \equiv \mathbf{D}$ for the diagonal matrix whose $i$-th diagonal element is $D_{i}$.   Notice that for any fixed $n$, the traditional and celebrated Fay-Herriot model of Section~\ref{HBayes} is a special case of the above $\Delta$-FH framework. Also, a triangular framework corresponding to Section~\ref{unstrBayes} can also be developed along the lines of $\Delta$-FH, however, the lack of finite-dimensional parameters makes the results on ranking either trivial or entirely dependent on technical assumptions, hence we omit the details. 

Consider the (generic) rank vector $\mathbf{R}_{n}$ at the $n$-th stage, which  is a permutation of $\mathbb{N}_{n} = \{ 1, 2, \ldots, {n} \}$. In keeping with standard mathematical conventions, let $\boldsymbol{\sigma}_{n} = (\sigma_{1}, \ldots, \sigma_{n})^{T}$ be a generic permutation of $\mathbb{N}_{n}$, and the set of all permutations of $\mathbb{N}_{n}$ is denoted by $\sigma(\mathbb{N}_{n})$. Thus, $\mathbf{R}_{n}$  takes values $\boldsymbol{\sigma}_{n}$ with associated probabilities, and our studies are investigating this probabilistic structure deeply in this  context. A major challenge is that with $n \to \infty$, the parameter space $\boldsymbol{\Theta}_{m_{n}}$ changes, the posterior distribution of $\boldsymbol{\theta}_{n} \in \boldsymbol{\Theta}_{m_{n}}$  given $\mathbf{Y}_{n}$ changes, and the support of the rank vectors $\mathbb{N}_{n}$ changes, and consequently the probabilistic properties of $\mathbf{R}_{n}$  also change in intricate ways. Our triangular array framework is designed to address these challenges.

For any pairs of positive integers $n_{1} < n_{2}$, we use the notations $\mathbf{Y}_{n_{2}}^{(n_{1})}$ ($\mathbf{D}_{n_{2}}^{(n_{1})}$, 
$\boldsymbol{\theta}_{n_{2}}^{(n_{1})}$, $\mathbf{R}_{n_{2}}^{(n_{1})}$ and so on) to denote the subvector of $\mathbf{Y}_{n_{2}}$ ($\mathbf{D}_{n_{2}}$, $\boldsymbol{\theta}_{n_{2}}$, $\mathbf{R}_{n_{2}}$ and so on) containing the first $n_{1}$ elements only. Our first result compares the posterior ranks and distributions of $\boldsymbol{\theta}_{n_{1}}$ and $\boldsymbol{\theta}_{n_{2}}^{(n_{1})}$, since these two random vectors have natural similarities both technically and in terms of their interpretations. The technical assumptions underpinning the results  are  in the supplementary materials, where we present the proofs also. In the special case where $\{ \boldsymbol{\beta}_{n} \equiv \boldsymbol{\beta} \}$ and $\{ A_{n} \equiv A \}$ in the $\Delta$-FH model, the proofs can be simplified slightly.

\begin{Theorem} \label{Thm:1}
    Under \textbf{Assumption-A}, 
    \begin{align}
      \sup_{\boldsymbol{\sigma}_{n_{1}} \in \sigma(\mathbb{N}_{n_{1}})}  \big| \mathbb{P} [ \mathbf{R}_{n_{1}} = \boldsymbol{\sigma}_{n_{1}} ] - \mathbb{P} [ \mathbf{R}_{n_{2}}^{(n_{1})} = \boldsymbol{\sigma}_{n_{1}} ]  \big| = O (n_{1}^{-1/2}).
        \label{eq:Thm1}
    \end{align}
almost surely.
Under the additional condition that $\mathbf{X}_{n_{1}} \boldsymbol{\beta}_{n_{1}} - \mathbf{X}_{n_{2}}^{(n_{1})} \boldsymbol{\beta}_{n_{2}} = O (n_{1}^{-1/2})$ almost surely, the Wasserstein-2 distance between the posterior distributions of  $n_{1}^{-1/2} \boldsymbol{\theta}_{n_{1}}$ and $n_{1}^{-1/2} \boldsymbol{\theta}_{n_{2}}^{(n_{1})}$ is  also $O (n_{1}^{-1/2})$ almost surely.

\end{Theorem}

The first part of Theorem~\ref{Thm:1} asserts that under very broad conditions, the posterior probabilities of the ranks of $\boldsymbol{\theta}_{n_{1}}$ and $\boldsymbol{\theta}_{n_{2}}^{(n_{1})}$ form a convergent Cauchy sequence almost surely. In  small area applications, the actual posterior distributions of $\boldsymbol{\theta}_{n_{1}}$ and $\boldsymbol{\theta}_{n_{2}}^{(n_{1})}$ also matter greatly. For example, for resource allocation, the ranking provides the rationale to allocate resources to some small areas and not others, while the actual posterior distribution of $\boldsymbol{\theta}_{n}$ may be used to decide on the amount of resource to provide to each selected small area.  The second result in Theorem~\ref{Thm:1} shows the Wasserstein distance between the distributions of $n_{1}^{-1/2} \boldsymbol{\theta}_{n_{1}}$ and $n_{1}^{-1/2} \boldsymbol{\theta}_{n_{2}}^{(n_{1})}$ also tends to zero under extremely broad and natural conditions. The additional technical assumption for the second part is easily satisfied, for example, by requiring that for every $n_{1} < n_{2}$, $\mathbf{X}_{n_{2}}^{(n_{1})} = \mathbf{X}_{n_{1}}$  and $\mathbf{Y}_{n_{2}}^{(n_{1})}  = \mathbf{Y}_{n_{1}}$ and some routine algebra.

While $\boldsymbol{\theta}_{n_{1}}$ and $\boldsymbol{\theta}_{n_{2}}^{(n_{1})}$ are comparable, in Theorem~\ref{Thm:1} we do not exploit or study the fact that for the latter, there is scope of additional borrowing strength, which may be of considerable importance. Consequently, we also present results below that compare $\boldsymbol{\theta}_{n_{1}}$ and $\boldsymbol{\theta}_{n_{2}}$, even though these two random vectors are of different dimensions. Recent advancements in probability theory (see \cite{ref:JAppliedProbability221178} and  \cite{ref:IEEE_IT224020_Wassterstein_DifferentSupport}) make such comparisons possible in the $\Delta$-FH framework. The framework of the latter requires comparing linear combinations of $\boldsymbol{\theta}_{n_{2}}$ with that of $\boldsymbol{\theta}_{n_{1}}$, which is not statistically meaningful in small area contexts. However, Theorem~\ref{Thm:1} may be viewed as an application of the  Cai-Lim framework \citep{ref:IEEE_IT224020_Wassterstein_DifferentSupport} when restricted to the context of ranking small areas. Below, we use the Gromov-Wasserstein distance to establish that the distance between the distributions of $n_{1}^{-1/2} \boldsymbol{\theta}_{n_{1}}$ and $n_{2}^{-1/2} \boldsymbol{\theta}_{n_{2}}$ is $O (n_{1}^{-1/2})$ almost surely as well. 

\begin{Theorem} \label{Thm:2}
    Under \textbf{Assumption-A} and \textbf{Assumption-B}, the   Gromov-Wasserstein distance between the posterior distributions of $n_{1}^{-1/2} \boldsymbol{\theta}_{n_{1}}$ and $n_{2}^{-1/2} \boldsymbol{\theta}_{n_{2}}$ is $O (n_{1}^{-1/2})$ almost surely.
  
\end{Theorem}

This result establishes that as $n_{1} \to \infty$ and $n_{1} < n_{2}$, the posterior distributions of $n_{1}^{-1/2} \boldsymbol{\theta}_{n_{1}}$ and $n_{2}^{-1/2} \boldsymbol{\theta}_{n_{2}}$ concentrate to a limit, and hence quantities derived from these (like rank vectors) would also have an asymptotic limit.

\section{Summary and conclusions} \label{Summary}

Small area statistics typically arise in contexts where there is a limited amount of directly observed information available for subpopulations, entities or subdomains of interest, and statistical analysis, prediction and inference must proceed by borrowing strengths across entities. There is broad gamut of topics that share similar characteristics, including applications ranging from personalized medicine to remote sensing. Rather than focusing solely on parameter estimation, several authors have investigated the identification of ``unusual'' populations, defined as those exhibiting extreme characteristics such as the highest poverty rates, lowest literacy levels, or substandard healthcare delivery \citep{morris1996hierarchical, GA_1999, HJ_2011, LP_2020}. For example, \citet{GA_1999} computed probabilities of relevant events to identify the worst-affected communities in environmental studies. \citet{HJ_2011} utilized a common mean hierarchical model, which is a special case of the model described in equation (\ref{eq:FH_Prior}) of Section~\ref{HBayes}, under an empirical Bayes approach (where model parameters $\beta$ and $\sigma^2$ are estimated from the data) to identify unusual healthcare providers. They evaluated predictive $p$-values and posterior probabilities to determine whether a provider's true mean exceeded an externally specified target or quantile. Similarly, in the context of small area estimation, \citet{LP_2020} considered the identification of areas whose true means (such as a poverty index) exceeded specific externally defined thresholds or estimated the worst poverty rate. Thus, in many applications, it is of interest to identify the best (or worst) ranking entities. While these existing approaches focus on threshold exceedance or local anomaly detection, our primary inferential objective centers on the comprehensive ranking of all populations or targeted subsets thereof. Our framework not only identifies the regions that have the lowest rank, it can also estimate the means or the smallest of the means of these regions.

 Naive ranking based on limited data or some estimator thereof lacks proper uncertainty quantification, and can be misleading. Following the seminal frequentist work of \cite{KWW2020}, in this paper we propose Bayesian approaches. While we study two priors and two approaches for obtaining credible sets, the principles we lay down here can be used if other priors or Bayesian inferential techniques are used. We demonstrate that the Bayesian approaches comprehensively outperform the non-Bayesian approaches of KWW and MRSW, and using a hierarchical Bayesian framework and an elliptical credible set seems the best strategy. Elliptical Bayesian credible sets are often orders of magnitude smaller than Cartesian ones or frequentist confidence sets that have the same level of coverage, while having similar or better properties in terms of the rank vectors and their associated weights. The \textit{interpretation} of the Bayesian results on ranking, which obtain relative weights  for all small areas, is interesting but challenging at first glance, since there are multiple sources of randomness and variability. A technically correct way of interpreting the empirical marginal relative weights given by $\mathbf{R}_{Bayes, j}$ for the $j$-th entity or small area is in terms of bounds of support, moments, and other properties of the rank that this entity may take. Using the set $\mathcal{S}_{Bayes, \alpha}$ to restrict the values of $\boldsymbol{\theta}$ results in computational efficiency, and robustness against distortion of the results due to outlying posterior samples. Other interpretations of our findings will be explored in future.

We develop a new approach for theoretical analysis of  a triangular sequence of Fay-Herriot models, called $\Delta$-FH framework. We show that the proposed Bayesian approaches have strong theoretical properties using this framework. The $\Delta$-FH framework and associated studies may be of independent interest in the broad area of small area statistics, and will be pursued in future. Interestingly, under this framework, it may be expected that posterior concentration results will imply that inferences on ranks based on $ ( 1- \alpha)^{-1}[\boldsymbol{\theta} \mathcal{I}_{\{ \boldsymbol{\theta} \in \mathcal{S}_{Bayes, \alpha}  \}} | \mathbf{Y}]$ will be consistent; however, this remains a conjecture to prove.


\bigskip\noindent {\large{\bf Disclosure Statement:}} No competing interests to declare from the authors.

\vspace{-.2in}
\section*{Supplementary Materials} 

In Section~{S.1} we present a step-by-step guidance to implement our proposed algorithm, with additional discussions, focusing on the case when a small number of small areas are of interest.  Additional tables and figures are presented in Section~S.2, the set of full conditional pdfs for the Gibbs sampling are in Section~S.3, details of the techniques used for comparing sets are in Section~S.4, size measures on elliptical sets are in Section~S.5, and Section~S.6 is on details on the theory, including proofs of the main results.





\bibliographystyle{chicago}  
\bibliography{RankingBib_Rev2}

@article{HJ_2011,
    author = {Jones, H. E. and  Spiegelhalter, D. J.}, 
    title ={The Identification of ''Unusual'' Health-Care Providers From a Hierarchical Model},
    journal = {The American Statistician},
   volume = {65},
 number = {3},
pages ={154-163},
year = {2011}
}

@article{GA_1999,
    author = {Gelman, A. and  Price, P. N.}, 
    title ={All maps of parameter estimates are misleading},
    journal = {Statistics in Medicine},
   volume = {18},
 number = {},
pages ={3221-3234},
year = {1999}
}

@article{LP_2020,
  author = {Lahiri, P. and Suntornchost, J.},
  year ={2020},
  title ={A general Bayesian approach to meet diﬀerent
inferential goals in poverty research for small areas},
 vol = {21},
 number ={4},
 pages = {237-253},
 journal ={Statistics in Transition (new series)},
 doi ={10.21307/stattrans-2020-040}
 }

@article{GoldsteinSpiegelhalter1996JRSSA,
 ISSN = {09641998, 1467985X},
 URL = {http://www.jstor.org/stable/2983325},
 author = {Harvey Goldstein and David J. Spiegelhalter},
 journal = {Journal of the Royal Statistical Society. Series A (Statistics in Society)},
 number = {3},
 pages = {385--443},
 publisher = {[Wiley, Royal Statistical Society]},
 title = {League Tables and Their Limitations: Statistical Issues in Comparisons of Institutional Performance},
 urldate = {2024-07-25},
 volume = {159},
 year = {1996}
}

@article{Mogstad2024,
    author = {Mogstad, Magne and Romano, Joseph P and Shaikh, Azeem M and Wilhelm, Daniel},
    title = {Inference for Ranks with Applications to Mobility across Neighbourhoods and Academic Achievement across Countries},
    journal = {The Review of Economic Studies},
    volume = {91},
    number = {1},
    pages = {476-518},
    year = {2024},
    issn = {0034-6527},
    doi = {10.1093/restud/rdad006},
    url = {https://doi.org/10.1093/restud/rdad006},
    eprint = {https://academic.oup.com/restud/article-pdf/91/1/476/56205143/rdad006.pdf},
}

@article{BergerDeely1988,
author = {James O. Berger and John Deely},
title = {{A Bayesian approach to ranking and selection of related means with alternatives to analysis-of-variance methodology}},
journal = {Journal of the American Statistical Association},
volume = {83},
number = {402},
pages = {364-373},
year = {1988},
publisher = {Taylor & Francis},
doi = {10.1080/01621459.1988.10478606}}

@article{AitkinLongford1986,
author={Aitkin, M. and Longford, N.},
year={1986},
title={{Statistical modeling issues in school effectiveness studies (with discussion)}}, 
journal={Journal of the Royal Statistical Society, (A)},
volume={149},
pages={1--43},
}

@article{BlandBratcher1968,
author = {R.P. Bland and T.L. Bratcher},
title ={{A Bayesian approach to the ranking of binomial probabilities}},
journal = {Journal of Applied Mathematics}, 
volume={16}, 
publisher = {SIAM},
pages ={843--850},
year ={1968},
}

@article{GovindarajuluHarvey1974, 
author = {Z. Govindarajulu and Charles Harvey},
title ={Bayesian procedures for ranking and selection problems},
journal = {Ann. Inst. Statist. Math.}, 
volume={26}, 
pages={35--53},
year={1974},
}

@article{LairdLouis1989,
author = {Nan M. Laird and Thomas A. Louis},
title ={{Empirical Bayes ranking methods}},
journal = {Journal of Educational Statistics},
volume = {14},
number = {1},
pages = {29-46},
year = {1989},
doi = {10.3102/10769986014001029},
}

@book{berger1985book,
title={Statistical Decision Theory and Bayesian Analysis},
author={Berger, J.O.},
	year={1985},
	publisher={Springer},
}

@article{KWW2020,
  author={Martin Klein and Tommy Wright and Jerzy Wieczorek},
  title={{A joint confidence region for an overall ranking of populations}},
  journal={Journal of the Royal Statistical Society Series C},
  year=2020,
  volume={69},
  number={3},
  pages={589-606},
  doi={10.1111/rssc.12402},
}

@article{Bechhofer1954,
author = {Robert E. Bechhofer},
title = {{A single-sample multiple decision procedure for ranking means of normal populations with known variances}},
volume = {25},
journal = {The Annals of Mathematical Statistics},
number = {1},
publisher = {Institute of Mathematical Statistics},
pages = {16 -- 39},
year = {1954},
doi = {10.1214/aoms/1177728845},
URL = {https://doi.org/10.1214/aoms/1177728845}
}

@phdthesis{gupta1989decision,
  title={On a decision rule for a problem in ranking means},
  author={Gupta, Shanti Swarup},
  year={1956},
  school={(and Mimeograph Series No. 150), Institute of Statistics, University Library, University of North Carolina at Chapel Hill}
}

@article{GoelRubin1977, 	
title={{On selecting a subset containing the best population-a Bayesian approach}},
author={Goel, P. and Rubin, H.},
year={1977}, 
journal={The Annals of Statistics}, 
volume={5},	
pages={969--983}
}

@article{EfronMorris1975, 
author={Bradley Efron and Carl Morris}, 
title={{Data analysis using Stein's estimator and its generalizations}},
journal={Journal of the American Statistical Association}, 
volume={70}, 
year={1975},
pages={311--319},
}

@article{fay1979estimates,
  title={{Estimates of income for small places: an application of James-Stein procedures to census data}},
  author={Fay, Robert E and Herriot, Roger A},
  journal={Journal of the American Statistical Association},
  volume={74},
  pages={269--277},
  year={1979},
  publisher={Taylor \& Francis}
}

@article{morris1996hierarchical,
  title={Hierarchical models for ranking and for identifying extremes, with applications},
 author={Morris, CN and Christiansen, CL},
  journal={Bayesian Statistics},
  volume={5},
  pages={277--296},
  year={1996},
  publisher={Oxford Univ. Press}
}

@book{rao2015small,
	title={Small area estimation},
	author={Rao, J. N. K. and Molina, Isabel},
	year={2015},
	publisher={Hoboken, NJ: John Wiley \& Sons, Inc.}
}

@article{ChungDatta2022,
title={Bayesian spatial models for estimating means of sampled and unsampled small areas},
author={Chung, H. and Datta, G.S.},
year={2022}, 
journal={Survey Methodology}, 
volume={48}, 
pages={463--489},
}

@article{ShenLouis1998,
title={Triple-goal estimates in two-stage hierarchical models},
author={Wei Shen and Thomas A. Louis},
journal={Journal of the Royal Statistical Society: Series B (Methodological)},
volume={60},
pages={455--471},
year={1998},
}

@article{Ghosh1996EstimationOM,
  title={{Estimation of median income of four-person families: a Bayesian time series approach}},
  author={Malay Ghosh and Narinder Nangia and Dal Ho Kim},
  journal={Journal of the American Statistical Association},
  year={1996},
  volume={91},
  pages={1423-1431},
  url={https://api.semanticscholar.org/CorpusID:121255867}
}

@article{hill1973diversity,
  title={Diversity and evenness: a unifying notation and its consequences},
  author={Hill, Mark O},
  journal={Ecology},
  volume={54},
  number={2},
  pages={427--432},
  year={1973},
  publisher={Wiley Online Library}
}

@article{ref:IEEE_IT224020_Wassterstein_DifferentSupport,
  title={Distances between probability distributions of different dimensions},
  author={Cai, Yuhang and Lim, Lek-Heng},
  journal={IEEE Transactions on Information Theory},
  volume={68},
  number={6},
  pages={4020--4031},
  year={2022},
  publisher={IEEE}
}

@article{ref:JAppliedProbability221178,
  title={Gromov--{W}asserstein distances between {G}aussian distributions},
  author={Delon, Julie and Desolneux, Agnes and Salmona, Antoine},
  journal={Journal of Applied Probability},
  volume={59},
  number={4},
  pages={1178--1198},
  year={2022},
  publisher={Cambridge University Press}
}

@misc{ref:Arxiv1810.08693v7_TotalVariation_Devroye,
  title={The total variation distance between high-dimensional {G}aussians with the same mean},
  author={Devroye, Luc and Mehrabian, Abbas and Reddad, Tommy},
  journal={arXiv preprint arXiv:1810.08693},
  year={2018}
}

@article{ref:AoS952145_GGS,
  title={On convergence of posterior distributions},
  author={Ghosal, Subhashis and Ghosh, Jayanta K and Samanta, Tapas},
  journal={The Annals of Statistics},
  volume={23},
  number={6},
  pages={2145--2152},
  year={1995},
  publisher={Institute of Mathematical Statistics}
}

@article{ref:AoS00500_GhosalGhoshvanderVaart_PosteriorConcentration,
  title={Convergence rates of posterior distributions},
  author={Ghosal, Subhashis and Ghosh, Jayanta K and van der Vaart, Aad W},
  journal={The Annals of Statistics},
  volume={28},
  number={2},
  pages={500--531},
  year={2000},
  publisher={Institute of Mathematical Statistics}
}

@article{ref:AoS07192_GhosalvanderVaart_PosteriorConcentration,
  title={Convergence Rates of Posterior Distributions for Noniid Observations},
  author={Ghosal, Subhashis and van der Vaart, Aad},
  journal={The Annals of Statistics},
  volume={35},
  number={1},
  pages={192--223},
  year={2007},
 publisher={Institute of Mathematical Statistics}
}

@book{jiang2007linear,
  title={Linear and generalized linear mixed models and their applications},
  author={Jiang, Jiming and Nguyen, Thuan},
  volume={},
  year = 2021,
  publisher={Springer}
}

@book{jiang2010large,
  title={Large sample techniques for statistics},
  author={Jiang, Jiming},
  year = 2010,
  publisher={Springer}
}

@inproceedings{heady2005eurarea,
  title={EURAREA: an overview of the project and its findings},
  author={Heady, Patrick and Ralphs, Martin},
  booktitle={SAE2005 Conference},
  pages={28--31},
  year={2005}
}

@misc{bell2003,
    title = {A {B}ayesian approach to recognizing Statistical Uncertainty
in ranked Survey Estimates},
    author = {Bell, W. R.},
    year = {2004},
    booktitle = {U.S. Census Bureau, Census Advisory Commit-
tee of Prefessional Association Meetings. April 22-23, 2004.}
    }

@misc{gomez2010bayesian,
  title={Bayesian statistics small area estimation},
  author={G{\'o}mez-Rubio, Virgilio and Best, Nicky and Richardson, Sylvia and Li, Guangquan and Clarke, Philip},
  year={2010},
  howpublished ={Imperial College London, \url{https://eprints.ncrm.ac.uk/id/eprint/1686/1/BayesianSAE.pdf}}
}

@techreport{wright2014ranking,
    author = {Wright, Tommy and Klein, Martin and Wieczorek, Jerzy},
    title = {Ranking populations based on sample survey data},
    institution = {Research Report Series, (Statistics 2014-12), Center for Statistical Research and Methodology, U.S. Bureau of the Census, \url{https://www.census.gov/library/working-papers/2014/adrm/rrs2014-12.html}},
    year = {2014}
}

@misc{wright2026estimated,
  title={Estimated Ranking Tables and Uncertainty},
  author={Wright, Tommy},
  year={2026},
  howpublished = {Preprint, available from \url{https://www.researchsquare.com/article/rs-9260965/latest}}
}

@article{willyard2025developing,
  title={Developing a data-driven system for identifying socially vulnerable populations and neighborhoods across the United States.},
  author={Willyard, Katherine Ann and Amaro, Gabriel and Sawyer, Chase and DeSalvo, Bethany and Basel, Wesley},
  journal={Journal of Emergency Management (Weston, Mass.)},
  volume={23},
  number={2},
  pages={125--136},
  year={2025}
}

@article{bijari2025method,
  title={A method for approximate rank confidence intervals in plant breeding experiments},
  author={Bijari, Reyhaneh and Vemireddy, Hanisha and Olafsson, Sigurdur},
  journal={Heliyon},
  volume={11},
  number={5},
  year={2025},
  publisher={Elsevier}
}

@article{moss2017comparing,
  title={Comparing percentages and ranks of adolescent weight-related outcomes among US states: implications for intervention development},
  author={Moss, Jennifer L and Liu, Benmei and Zhu, Li},
  journal={Preventive medicine},
  volume={105},
  pages={109--115},
  year={2017},
  publisher={Elsevier}
}

@article{waldrop2017ranking,
  title={Ranking states on coverage of cancer-preventing vaccines among adolescents: the influence of imprecision},
  author={Waldrop, Anne R and Moss, Jennifer L and Liu, Benmei and Zhu, Li},
  journal={Public Health Reports},
  volume={132},
  number={6},
  pages={627--636},
  year={2017},
  publisher={SAGE Publications Sage CA: Los Angeles, CA}
}

@article{al2022simultaneous,
  title={Simultaneous confidence intervals for ranks with application to ranking institutions},
  author={Al Mohamad, Diaa and Goeman, Jelle J and van Zwet, Erik W},
  journal={Biometrics},
  volume={78},
  number={1},
  pages={238--247},
  year={2022},
  publisher={Oxford University Press}
}

@article{chatzi2024prediction,
  title={Prediction-powered ranking of large language models},
  author={Chatzi, Ivi and Straitouri, Eleni and Thejaswi, Suhas and Rodriguez, Manuel G},
  journal={Advances in Neural Information Processing Systems},
  volume={37},
  pages={113096--113133},
  year={2024}
}

@inproceedings{chen2025ranking,
  title={Ranking-aware uncertainty for text-guided image retrieval},
  author={Chen, Junyang and Lai, Hanjiang},
  booktitle={International Conference on Intelligent Computing},
  pages={394--406},
  year={2025},
  organization={Springer}
}

@inproceedings{neuhof2024confident,
  title={Confident feature ranking},
  author={Neuhof, Bitya and Benjamini, Yuval},
  booktitle={International Conference on Artificial Intelligence and Statistics},
  pages={1468--1476},
  year={2024},
  organization={PMLR}
}

@article{rising2021uncertainty,
  title={Uncertainty in ranking},
  author={Rising, Justin},
  journal={arXiv preprint arXiv:2107.03459},
  year={2021}
}

@article{liao2026distribution,
  title={Distribution-informed Efficient Conformal Prediction for Full Ranking},
  author={Liao, Wenbo and Huang, Huipeng and Jia, Chen and Xi, Huajun and Zeng, Hao and Wei, Hongxin},
  journal={arXiv preprint arXiv:2601.23128},
  year={2026}
}

@article{aminghafouri2024joint,
  title={Joint confidence region approach to ranking hotspot locations considering uncertainty in expected risk estimates},
  author={Aminghafouri, Reza and Fu, Liping},
  journal={Transportation research record},
  volume={2678},
  number={1},
  pages={555--567},
  year={2024},
  publisher={SAGE Publications Sage CA: Los Angeles, CA}
}

@article{liu2021simple,
  title={Simple measures of uncertainty for model selection},
  author={Liu, Xiaohui and Li, Yuanyuan and Jiang, Jiming},
  journal={Test},
  volume={30},
  number={3},
  pages={673--692},
  year={2021},
  publisher={Springer}
}

@article{li2024measures,
  title={Measures of uncertainty for shrinkage model selection},
  author={Li, Yuanyuan and Jiang, Jiming},
  journal={Statistica Sinica},
  volume={34},
  number={1},
  pages={505--521},
  year={2024},
  publisher={JSTOR}
}

\clearpage


\begin{center}
    {\Large {\bf Credible Sets of Overall Ranking of Entities}}

\bigskip \bigskip

    {\Large {\bf Supplementary Materials}}

\end{center}

\setcounter{section}{0}
\setcounter{table}{0}
\setcounter{figure}{0}
\setcounter{equation}{0}
\setcounter{page}{1}
\renewcommand{\thesection}{S.\arabic{section}}
\renewcommand{\thetable}{S.\arabic{table}}
\renewcommand{\thefigure}{S.\arabic{figure}}
\renewcommand{\theequation}{S.\arabic{equation}}
\renewcommand{\thepage}{S.\arabic{page}}


\section{Implementation steps when a few small areas are of interest}
\label{Sec:Small_m}

Here, we provide a detailed layout of the steps one can use to reproduce our results, concentrating on the case where tables and displays with a small number of small areas is fo interest. 

\begin{figure}[b]\caption{Barplot of ranks of six mid-Atlantic states, when using  $\mathcal{R}_{HB, E, 0.1}$
		for commute time data. This figure corresponds to Table~\ref{Tab:Rank_dist}.} \label{Fig:RankComb} 
	\begin{center}
		\begin{tabular}{c}
			\includegraphics[scale=.95]{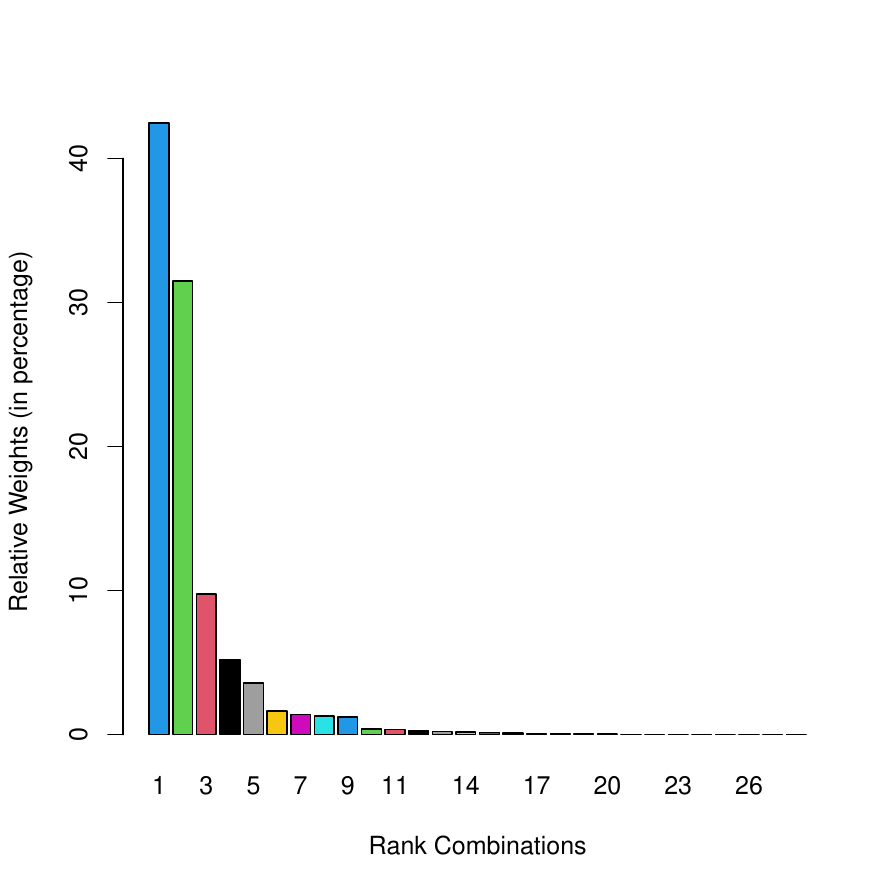}		
		\end{tabular}
	\end{center}
\end{figure}

\begin{enumerate}
    \item \textbf{Fix constants:} Fix the parameter $\alpha \in [0, 1)$. For the case of $\alpha = 0$, skip the next step and move directly to Step-\#3 below. 

    \item \textbf{Construct credible sets:} Fix the prior, and the kind of credible set of interest. The former can be unstructured Bayesian (UB) or heirarchical Bayesian (HB), the latter can be Cartesian (C) or elliptical (E). The resulting credible sets are denoted by $ \mathcal{S}_{UB, C, \alpha}$, $ \mathcal{S}_{HB, C, \alpha}$, $ \mathcal{S}_{UB, E, \alpha}$, or $ \mathcal{S}_{HB, E, \alpha}$. We use the generic notation $ \mathcal{S}_{Bayes, \alpha}$ to denote any of the above four kinds of sets. 

    \item \textbf{Rank:} Rank the elements of the credible set $\mathcal{S}_{Bayes, \alpha}$. In case $\alpha = 0$, rank all the vectors produced in the MC or MCMC step. The output will be a permutation of $\{1, 2, \ldots, m \}$. 

    \item \textbf{Construct a doubly stochastic matrix:} Construct the $m \times m$ doubly-stochastic matrix using the procedure laid out in Section~\ref{jt-cred-overall-rank}. Starting with an empty matrix,  for any  $\boldsymbol{\theta}^{(s)}$ in $\mathcal{S}_{Bayes, \alpha}$, if the $j$-th entry (i.e., small area $j$) has the $i$-th rank, we  add one in the $(i, j)$-th slot, for $j = 1, 2, \ldots, m$. Thus we get a matrix where every row and every column has exactly $(m-1)$ entries as zero and exactly one entry as one, consequently the matrix is \textit{a fortiori} doubly stochastic. The process is repeated for all elements of $\mathcal{S}_{Bayes, \alpha}$. Construct $\mathbf{R}_{Bayes}$ is the average of these matrices,  so it remains doubly stochastic.

    \item \textbf{Extract submatrices of interest:} Fix a small number of small areas, say $m_{0}$ where typically $m_{0} \ll m$ to analyze closely. Suppose the range of rank values that these $m_{0}$ small areas can take are $\tilde{m}$, where $\tilde{m} \leq m$. Extract the $\tilde{m} \times m_{0}$ submatrix from the above doubly stochastic matrix for further detailed study. An example of such submatrix is given in Table~\ref{Tab:Rank_dist}.
    
    These tables are respectively based on using the MCMC from the hierarchical Bayesian approach and elliptical credible set of 90\% coverage and empirical posterior distribution of rank vector based on entire MCMC sample in the HB framework for the states ID, KS, IA, AK, MT, NE, WY, ND, SD (states with lowest commute times) and on using the MCMC from the hierarchical Bayesian approach and elliptical credible set of 90\% coverage on MD, NY, NJ, DC, VA, PA (mid-Atlantic states, which have very high commute times). 

    \item \textbf{Interpretation:} When $\alpha = 0$ as in the case Table~\ref{tab:HB.9States}, the interpretation is straightforward: the table entries are the posterior probabilities (in percentages, for better display) corresponding to vaious states and ranks. For example, based on Table~\ref{tab:HB.9States}, we can claim that KS takes the ranks 4--9 with posterior probabilities  0.0002, 0.0098, 0. 0665, 0.3082, 0. 6102 and 0.0051 respectively. Similarly, the rank 6 is taken by KS, IA, AK, MT, NE, WY with posterior probabilities 0.0665, 0.1220, 0.3838, 0.1945, 0.0687 and 0.1644 respectively. 

    When $\alpha \in (0, 1)$, as in the case of Table~\ref{tab:HB.HPD.9States_251021} or Table~\ref{Tab:Rank_dist}, some additional care in needed for interpretation. For example, from Table~\ref{tab:HB.HPD.9States_251021}, we can claim that the column corresponding to AK shows six ranks 3-8, with relative weights 0.054, 0.110, 0.258, 0.539, 0.028, 0.010 
among the $90\%$ of the $\boldsymbol{\theta}$ vectors obtained using MCMC on the hierarchical Bayesian approach and elliptical credible set. In the same setup, row 6 assigns rank 6 to KS, IA, AK, MT, NE and WY with relative weights 0.019, 0.045, 0.539, 0.197, 0.039, 0.160. Note the difference of this case with the $\alpha = 0$ scenario: while the numeric values are comparable, this case is based only on the $90\%$ of the $\boldsymbol{\theta}$ vectors obtained using MCMC on the hierarchical Bayesian approach and elliptical credible set. Thus, we can use the table entries as relative weights. In this case, we can obtain upper and lower bounds for various posterior moments and other summaries, as detailed in the main paper.

    \item \textbf{Other summaries and displays:} The submatrices of interest may be presented in other formats as well, depending on the use-case and context. For Table\ref{Tab:Rank_dist} on the 6 states with some of the highest commute times, the 90\% hierarchical Bayesian elliptical credible set had only 28 combinations of ranks. These are also graphically displayed in Figure~\ref{Fig:RankComb}.  

    As an alternative to selecting the \textit{states} to focus on, we may concentrate on a subset of ranks instead. For example if we 
focus on the ranks $\{ 47, 48, 49, 50, 51 \}$, that is, the states with the longest commute times, then there are only a few states that appear, and Table~\ref{Tab:Top_Ranks} contains this information. The state-combinations (MD, NY, NJ, DC, IL), (MD, NY, NJ, DC, MA) and (MD, NY, NJ, DC, VA) with longest five commute times  with respective weights 88.11, 11.84 and 0.04, within the the 90\% hierarchical Bayesian elliptical credible set.

\end{enumerate}

While we discuss the case of $\alpha = 0$ both in the implementation above as well as in the main paper, we note some caveats here. 
 First, skipping the credible set construction step (thus effectively setting $\alpha = 0$), results in inclusion of all MC (or MCMC) samples, some of which may be extreme outliers and present in very low posterior probability regions. This results in an obvious practical robustness problem, since the rank vector from such a $\boldsymbol{\theta}$ may not reflect its extreme nature at all. 
 
 To appreciate this, note that it is overwhelmingly likely that rank vectors based on  MCMC samples are all unique, and provide no useful summarization at all. To appreciate this, consider the commute data  Example~\ref{Exmp:1}. The number of possible rank vectors for the 51 small areas  is $51! = O (10^{66})$, about 2.5 times the size of the observed universe in meters. Even an extremely high choice of MCMC sample size, say $10^{10}$, would result in each realized rank vector to have vanishingly small probability $O (10^{-56})$. Consequently, meaningful inference using the empirical rank vectors from the MCMC samples involving all the small areas is impossible.

    The problem is exacerbated if we upend the importance of $\boldsymbol{\theta}$ vectors and their ranks, and treat the latter as the primary objects of interest. Note that in real data applications, one would typically be interested in prediction, inference and uncertainty quantification of $\boldsymbol{\theta}$ conditional on the data as these directly reflect on stakeholder interests and inform policy. Thus, credible sets like $\mathcal{S}_{Bayes, \alpha}$ for $\alpha \in (0, 1)$ are required anyway for decision making. The ranks and associated relative weights of the small areas are of importance in ensuring fairness, in making sure that chance variations do not significantly impact fund allocations and downstream projects and other tasks, and hence we study this in our paper, but a focus on the ranks should not distract us from the fact that these are coupled with actual $\boldsymbol{\theta}$ values that inform the actual policies. If one sets $\alpha = 0$, obtains ranks, performs some kind of inference like obtaining a credible set of ranks, and then traces back the $\boldsymbol{\theta}$ that corresponds to a rank vector within that credible set: there is no guarantee that such an elicited $\boldsymbol{\theta}$ will be in some acceptable $\mathcal{S}_{Bayes, \alpha}$, precisely because of the problem described above. With such an approach, one has to separately obtain credible sets for the ranks and for $\mathcal{S}_{Bayes, \alpha}$ and maintain a registry of records linking ranks to $\mathcal{S}_{Bayes, \alpha}$ values, thus overwhelmingly negating any computational efficiency of starting with $\alpha = 0$ in the first place. Even then,  having to regularly look up the registry linking ranks to values to ensure that a given rank (value) corresponds to a value (rank) within a specified credible set is a major ask for downstream tasks. 

Nevertheless, one must recognize the strengths of using $\alpha = 0$ also. Unlike the case for $\alpha \in (0, 1)$ presented in Section~\ref{Sec:Interpretation}, posterior moments or probabilities are straightforward computations and do not involve inequalities, and are thus more intuitive and have simpler interpretations. 

\clearpage
\section{Additional tables and figures}
\label{Sec:Supple_Tables}

\begin{table}[h]\caption{Commute times: Number of rank vectors included in the 90\% confidence/credible set according to different methods when  $k$ states are considered, for $k = 4, \ldots, 10$, ordered in terms of increasing average commute times. The states are included in the order of their ranks (for $y_i$): SD, ND, WY, NE, MT, AK, IA, KS, ID, OK. } \label{Tab:Supple_NumerOfRankVectors}
\begin{center}
\begin{tabular}{|c|rrrrrr|}
  \hline
 Number of States & KWW & MRSW & UB.Cart & UB.Ellip & HB.Cart & HB.Ellip \\ 
  \hline
 4 &  14 &   8 &   8 &   8 &   8 &   8 \\ 
 5 &  78 &  20 &  22 &  22 &  34 &  32 \\ 
 6 & 336 &  72 &  74 &  75 & 105 & 100 \\ 
 7 & 1320 & 228 & 141 & 141 & 168 & 152 \\ 
 8 & 12576 & 960 & 395 & 371 & 430 & 382 \\ 
 9 & 43008 & 960 & 423 & 389 & 471 & 455 \\ 
 10 & 43008 & 1570 & 407 & 398 & 453 & 434 \\ 
   \hline
\end{tabular}
\end{center}
\end{table}

\begin{table}\caption{Additional details on the states with lowest commuting times (SD, ND, WY, NE). Here we show the 8 rank vectors contained within the Bayesian credible sets, with their respective weights (in percentages). This table is an expansion of the Bayesian portion of the top row of Table~\ref{Tab:Supple_NumerOfRankVectors}. } \label{Table:Supple_4States_New}
 
\begin{center}
  \begin{tabular}{|ll|cccccccc|}
    \hline
\multicolumn{2}{|r}{Rank orders}   &  3421 &  4321 &  4312 &  3412 & 4231 & 4213 & 4123 & 4132 \\
     \hline
 \multirow{2}*{UB} & Cartesian &    25.45 & 24.69 & 24.29 & 23.59 & 1.19 & 0.69 & 0.07 & 0.04\\
                   & Elliptical&    25.46 & 24.70 & 24.36 & 23.58 & 1.10 & 0.61 & 0.08 & 0.12\\
\hline
 \multirow{2}*{HB} & Cartesian &    20.92 & 29.11 & 27.10 & 18.44 & 2.38 & 1.54 & 0.24 & 0.27\\
                   & Elliptical&    20.82 & 29.35 & 27.22 & 18.35 & 2.32 & 1.40 & 0.21 & 0.32\\
    \hline
  \end{tabular}
\end{center} 
\end{table}

\begin{table}[h]\caption{Subset of states scoring the top 5 ranks, when using  $\mathcal{R}_{HB, E, 0.1}$.}\label{Tab:Top_Ranks}
\begin{center}
{\scriptsize
\begin{tabular}{cccccr}
  \hline
  \multicolumn{5}{c}{Ranks} &  Frequency \\
   51 & 50 & 49 & 48 & 47 & (in \%) \\ 
  \hline
   MD & NY & NJ & DC & IL & 88.11 \\ 
   MD & NY & NJ & DC & MA & 11.84 \\ 
   MD & NY & NJ & DC & VA & 0.04 \\ 
   \hline
\end{tabular}
}    
\end{center}
\end{table}

\begin{table}[h]\caption{Ranks of six mid-Atlantic states, when using  $\mathcal{R}_{HB, E, 0.1}$. Apart from the 28 rank combinations reported below, the other {$ {51 \choose 6} \times 6!- 28 \approx 1.3 \times 10^{10}$} rank combinations did not appear within the 90\% credible elliptical set.}\label{Tab:Rank_dist}
\begin{center}
{\scriptsize
\begin{tabular}{|c|cccccc|c|}
  \hline
Rank & \multicolumn{6}{c}{Ranks of States} & Rel. weights \\
Combination &  MD & NY & NJ & DC & VA & PA & (in \%) \\
  \hline
 1 &   51 & 50 & 49 & 48 & 45 & 41 & 42.46 \\
 2 &   51 & 50 & 49 & 48 & 45 & 40 & 31.50 \\
 3 &   51 & 50 & 49 & 48 & 45 & 39 & 9.74 \\
 4 &   51 & 50 & 48 & 49 & 45 & 41 & 5.20 \\
 5 &   51 & 50 & 48 & 49 & 45 & 40 & 3.58 \\
 6 &   51 & 50 & 49 & 48 & 46 & 41 & 1.63 \\
 7 &   51 & 50 & 49 & 48 & 45 & 38 & 1.39 \\
 8 &   51 & 50 & 48 & 49 & 45 & 39 & 1.28 \\
 9 &   51 & 50 & 49 & 48 & 46 & 40 & 1.22 \\
10 &   51 & 50 & 49 & 48 & 44 & 41 & 0.38 \\
11 &   51 & 50 & 49 & 48 & 46 & 39 & 0.36 \\
12 &   51 & 50 & 49 & 48 & 44 & 40 & 0.28 \\
13 &   51 & 50 & 48 & 49 & 46 & 41 & 0.21 \\
14 &   51 & 50 & 48 & 49 & 45 & 38 & 0.17 \\
15 &   51 & 50 & 48 & 49 & 46 & 40 & 0.13 \\
16 &   51 & 50 & 49 & 48 & 44 & 39 & 0.10 \\
17 &   51 & 50 & 48 & 49 & 44 & 41 & 0.08 \\
18 &   51 & 50 & 48 & 49 & 46 & 39 & 0.08 \\
19 &   51 & 50 & 49 & 48 & 45 & 37 & 0.06 \\
20 &   51 & 50 & 49 & 48 & 44 & 38 & 0.03 \\
21 &   51 & 50 & 49 & 48 & 45 & 42 & 0.02 \\
22 &   51 & 50 & 49 & 48 & 47 & 40 & 0.02 \\
23 &   51 & 50 & 48 & 49 & 44 & 40 & 0.02 \\
24 &   51 & 50 & 49 & 48 & 47 & 41 & 0.02 \\
25 &   51 & 50 & 49 & 48 & 46 & 37 & 0.01 \\
26 &   51 & 50 & 49 & 48 & 46 & 38 & 0.01 \\
27 &   51 & 50 & 48 & 49 & 44 & 39 & 0.01 \\
28 &   51 & 50 & 48 & 49 & 45 & 37 & 0.01 \\
   \hline
\end{tabular}
}
\end{center}
\end{table}

\clearpage
\begin{figure}\caption{The 90\% joint confidence region for commute time ranking using MRSW approach, with  $\mathcal{R}_{UB, C, \alpha}$-based Bayesian ranks with associated weights overlay. The corresponding figure that replicates the KWW approach is contained in Figure~\ref{Fig:Commute} of the main paper.}\label{Fig:Commute_SP}
\begin{center}
    \includegraphics[width=.49\textwidth,angle=90]{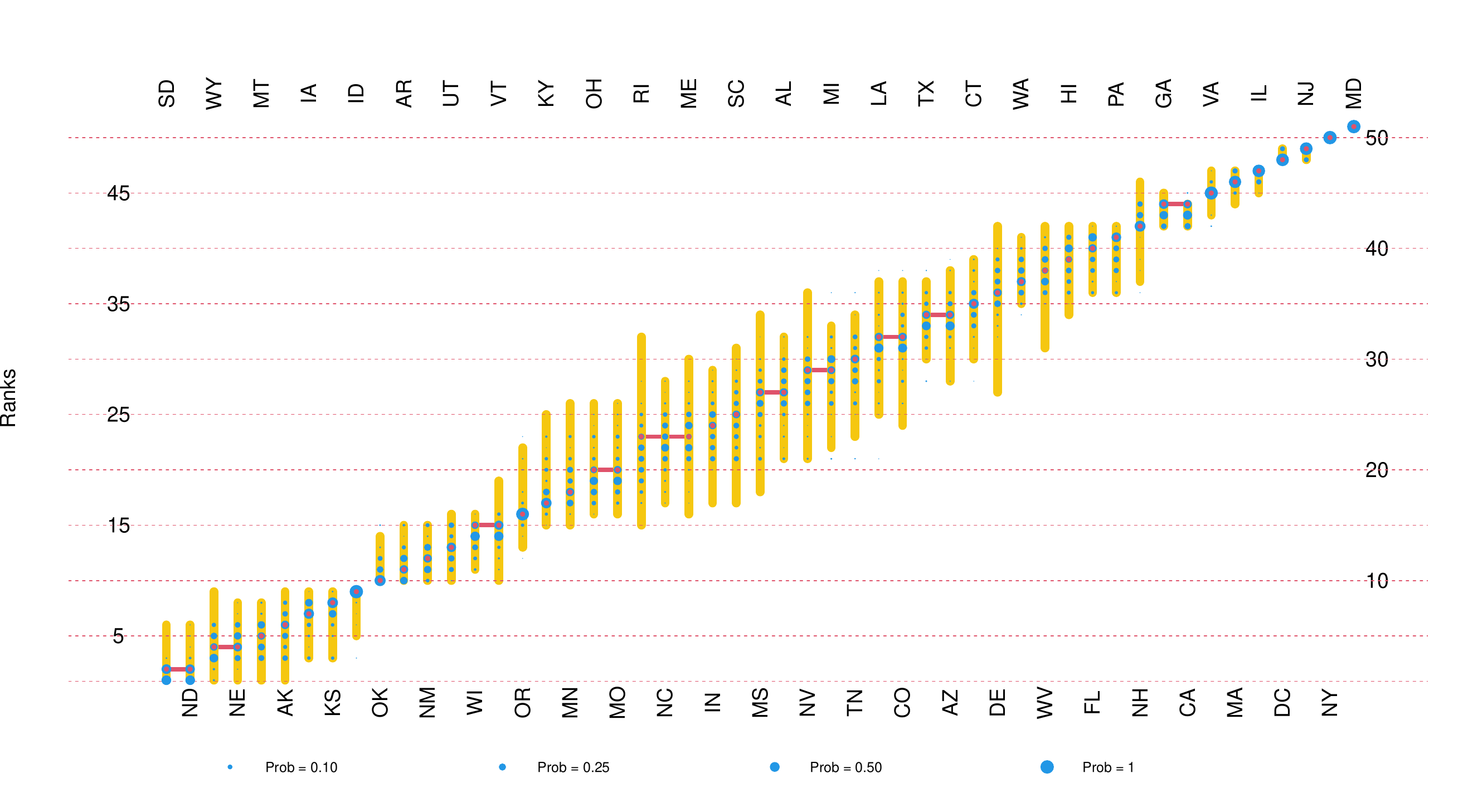}  
\end{center}
\end{figure}
The Figure~\ref{Fig:Commute_SP} has the 90\% confidence region using the MRSW approach. As is evident, there is little difference between the outputs of KWW and MRSW, with both these non-Bayesian techniques typically resulting in wider bands of ranks for most US states. The bands from MRSW are sometimes marginally  narrower than those from KWW, but nevertheless are non-competitive with Bayesian approaches.
This can also be seen from Table~\ref{Tab:Supple_NumerOfRankVectors}, where we considered the $k$  states with lowest commute times sequentially, for $k = 4, \ldots, 10$ and counted the number of rank vectors that are included in the 90\% confidence/credible sets  
$\mathcal{R}_{KWW, 0.1}$, $\mathcal{R}_{MRSW, 0.1}$, $\mathcal{R}_{UB, C, 0.1}$, $\mathcal{R}_{UB, E, 0.1}$, $\mathcal{R}_{HB, C, 0.1}$, $\mathcal{R}_{HB, E, 0.1}$. Notice that the both $\mathcal{R}_{KWW, 0.1}$ and $\mathcal{R}_{MRSW, 0.1}$ rapidly grow in size compared to the credible sets, the former growing extremely fast. This suggests that even for applications involving moderate number of small areas, using the non-Bayesian approaches may not be computationally viable.


\begin{figure}[h]\caption{The  90\% joint confidence region for income ranking using the MRSW approach, with  $\mathcal{R}_{HB, E, 0.1}$-based Bayesian ranks with associated weights overlay. The corresponding figure using the KWW approach is presented in Figure~\ref{Fig:Income} of the main paper.}\label{Fig:Income_SP}
\begin{center}
    \includegraphics[width=.97\textwidth]{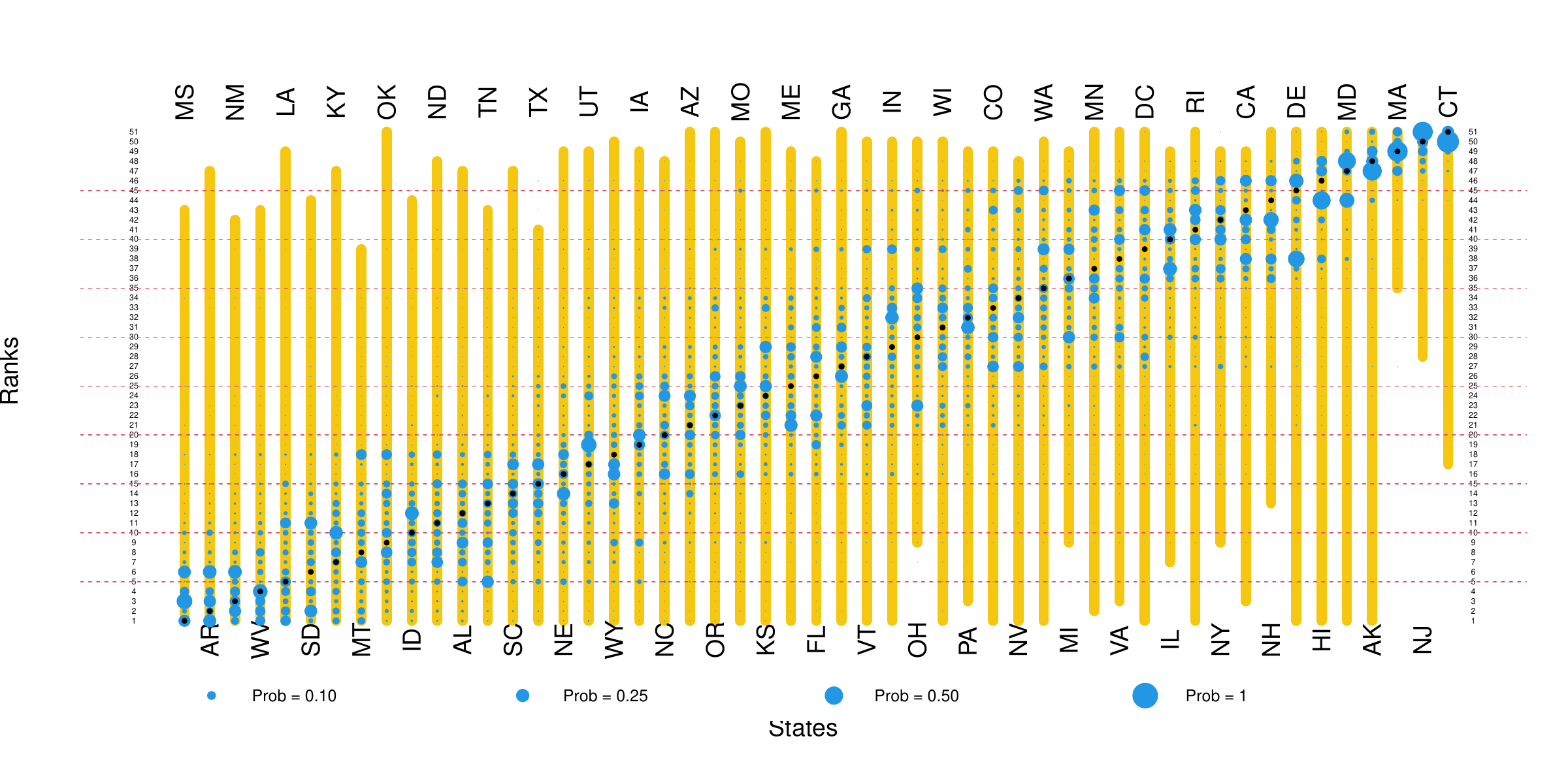}  
\end{center}
\end{figure}

\begin{table}[h]\caption{Mean absolute deviation (${AMD}$) for the  median income example from Section~\ref{medianincome} (Part-1).}
\begin{center}
{\scriptsize
\begin{tabular}{|c|rrrrrr|}
  \hline
States & KWW & MRSW & UB-Cart & UB-Ellip & HB-Cart & HB-Ellip \\ 
  \hline
  MS & 22.50 & 21.00 & 5.90 & 4.97 & 3.87 & 2.63 \\ 
  AR & 22.54 & 22.04 & 8.81 & 7.74 & 5.34 & 3.04 \\ 
  NM & 20.13 & 18.64 & 3.02 & 2.20 & 2.58 & 1.54 \\ 
  WV & 19.76 & 18.28 & 4.10 & 3.06 & 2.94 & 1.46 \\ 
  LA & 20.41 & 20.41 & 12.92 & 12.00 & 8.52 & 5.74 \\ 
  SD & 18.64 & 17.18 & 4.29 & 3.89 & 3.08 & 3.41 \\ 
  KY & 18.38 & 17.89 & 4.47 & 4.24 & 4.43 & 3.22 \\ 
  MT & 15.77 & 13.44 & 4.65 & 4.64 & 3.82 & 2.83 \\ 
  OK & 18.41 & 18.41 & 19.36 & 19.93 & 7.51 & 4.61 \\ 
  ID & 15.91 & 14.55 & 3.92 & 2.97 & 4.12 & 2.86 \\ 
  ND & 16.24 & 15.79 & 8.48 & 7.79 & 4.38 & 3.60 \\ 
  AL & 15.25 & 14.81 & 5.19 & 4.06 & 4.23 & 2.75 \\ 
  TN & 14.32 & 12.63 & 6.66 & 6.87 & 4.77 & 3.07 \\ 
  SC & 14.29 & 13.87 & 6.93 & 6.12 & 4.81 & 3.18 \\ 
  TX & 12.27 & 11.12 & 5.92 & 6.26 & 5.39 & 3.08 \\ 
  NE & 14.71 & 13.90 & 13.93 & 14.61 & 8.04 & 3.97 \\ 
  UT & 13.94 & 13.55 & 7.42 & 6.63 & 5.10 & 2.50 \\ 
  WY & 14.00 & 13.62 & 8.98 & 7.94 & 6.10 & 6.16 \\ 
  IA & 12.98 & 12.98 & 6.10 & 5.85 & 4.71 & 2.72 \\ 
  NC & 12.42 & 12.42 & 3.22 & 2.85 & 3.55 & 2.07 \\ 
  AZ & 13.24 & 13.24 & 10.26 & 8.41 & 6.85 & 4.43 \\ 
  OR & 13.06 & 13.06 & 10.30 & 9.56 & 6.39 & 3.28 \\ 
  MO & 12.92 & 12.62 & 7.76 & 6.63 & 6.02 & 4.41 \\ 
  KS & 12.82 & 12.82 & 11.51 & 12.19 & 5.82 & 4.04 \\ 
  ME & 12.50 & 12.24 & 7.41 & 6.17 & 6.09 & 3.57 \\ 
  FL & 12.04 & 12.04 & 6.57 & 7.04 & 5.22 & 3.16 \\ 
   \hline
\end{tabular}
}
\end{center}
\end{table}

\begin{table}\caption{Mean absolute deviation (${AMD}$) for the  median income example from Section~\ref{medianincome} (Part-2).}
\begin{center}
{\scriptsize
\begin{tabular}{|c|rrrrrr|}
  \hline
 States & KWW & MRSW & UB-Cart & UB-Ellip & HB-Cart & HB-Ellip \\ 
  \hline 
  GA & 12.76 & 12.76 & 14.03 & 15.11 & 8.16 & 6.06 \\ 
  VT & 12.82 & 12.62 & 6.97 & 5.84 & 5.43 & 3.73 \\ 
  IN & 12.92 & 12.74 & 6.61 & 6.37 & 5.47 & 3.60 \\ 
  OH & 12.12 & 10.50 & 9.19 & 9.49 & 7.72 & 5.58 \\ 
  WI & 13.24 & 13.10 & 5.71 & 4.22 & 4.78 & 3.22 \\ 
  PA & 13.24 & 12.51 & 4.22 & 3.64 & 3.70 & 2.33 \\ 
  CO & 13.55 & 13.55 & 14.37 & 13.92 & 7.84 & 4.50 \\ 
  NV & 13.90 & 13.88 & 20.36 & 21.60 & 10.08 & 4.16 \\ 
  WA & 14.33 & 14.30 & 5.50 & 4.99 & 4.66 & 3.15 \\ 
  MI & 13.90 & 11.44 & 3.63 & 3.43 & 3.90 & 3.22 \\ 
  MN & 15.12 & 14.70 & 5.98 & 5.45 & 4.77 & 3.68 \\ 
  VA & 15.57 & 14.71 & 8.03 & 8.01 & 6.16 & 6.09 \\ 
  DC & 16.06 & 16.06 & 20.23 & 23.07 & 9.54 & 6.77 \\ 
  IL & 16.70 & 14.09 & 5.81 & 5.43 & 3.63 & 1.96 \\ 
  RI & 17.16 & 17.16 & 5.54 & 4.31 & 4.34 & 2.51 \\ 
  NY & 15.60 & 14.37 & 5.05 & 4.05 & 2.56 & 1.73 \\ 
  CA & 18.86 & 17.89 & 12.40 & 12.17 & 7.87 & 4.76 \\ 
  NH & 15.30 & 13.44 & 4.22 & 4.13 & 2.82 & 2.26 \\ 
  DE & 19.82 & 19.82 & 18.05 & 18.93 & 10.70 & 8.45 \\ 
  HI & 20.59 & 20.59 & 8.41 & 5.67 & 4.99 & 3.22 \\ 
  MD & 21.39 & 21.39 & 4.60 & 2.39 & 1.31 & 1.13 \\ 
  AK & 22.24 & 22.24 & 17.09 & 16.68 & 3.11 & 1.16 \\ 
  MA & 10.73 & 6.35 & 1.26 & 1.25 & 1.02 & 0.46 \\ 
  NJ & 13.57 & 10.58 & 1.68 & 1.41 & 0.76 & 0.83 \\ 
  CT & 18.00 & 17.00 & 1.79 & 1.46 & 1.29 & 1.07 \\ 
  \hline
  Average & 15.67 & 14.91 & 8.02 & 7.60 & 5.10 & 3.39 \\ 
   \hline
\end{tabular}
}
\end{center}
\end{table}

\begin{figure}[b]\caption{
		For commute time data: Double scatter plots of joint HB posterior samples for ID and SD vs. AK. Note that $\mbox{ID} \approx > ~\mbox{AK }\approx > \mbox{SD}$.        } \label{FigS1:ID_AK_SD} 
	\begin{center}
		\begin{tabular}{c}
			\includegraphics[scale=.5]{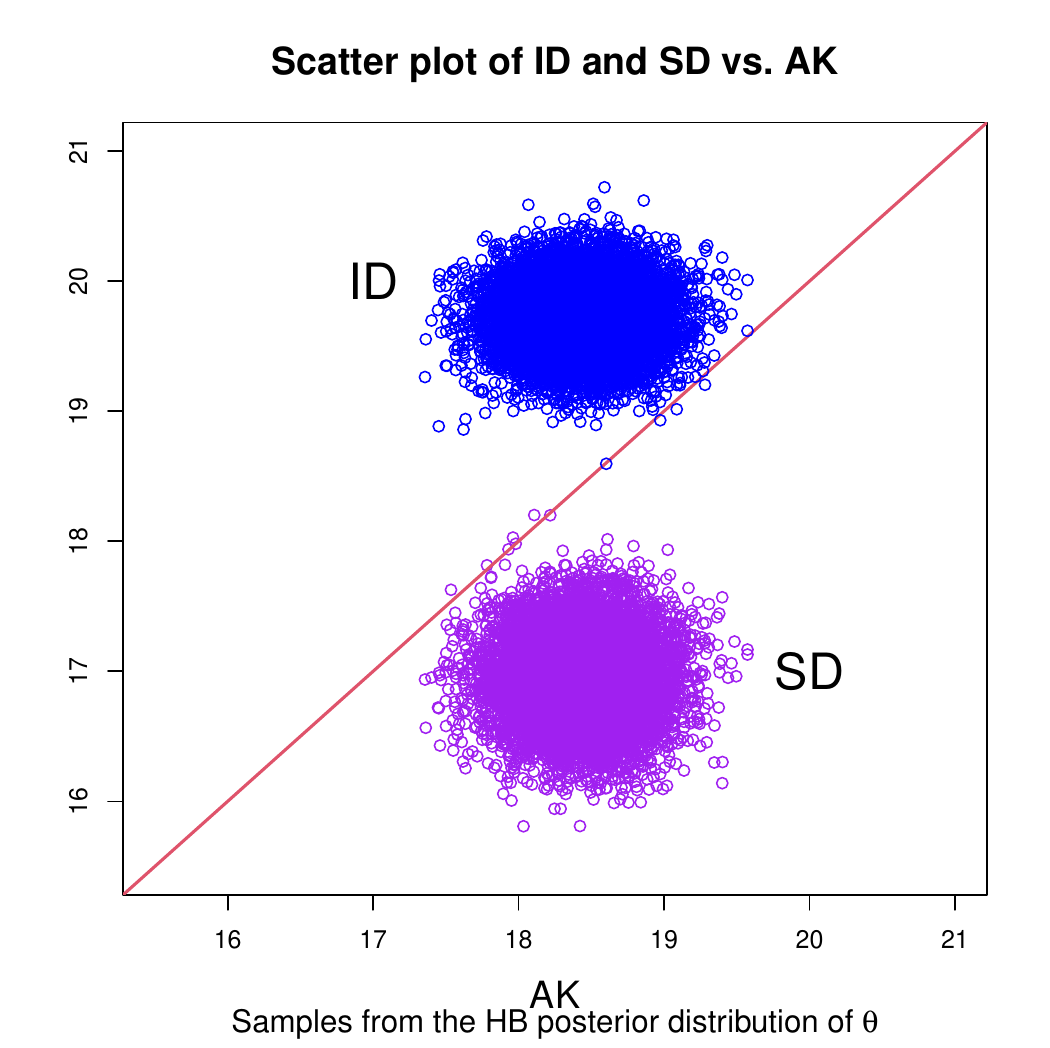}		
		\end{tabular}
	\end{center}
\end{figure}

\clearpage

\section{The set of full conditionals for Gibbs sampling } \label{full-condnls}

The joint posterior pdf of $\boldsymbol{\theta}, \boldsymbol{\beta}, A$ is 
 \begin{eqnarray*}
   \pi(\boldsymbol{\theta}, \boldsymbol{\beta}, A|\mathbf{y}) &\propto& \exp[-\frac 12 (\mathbf{y}-\boldsymbol{\theta})^T \mathbf{D}^{-1}(\mathbf{y} -\boldsymbol{\theta})] \times A^{-\frac m2} \exp[-\frac 1{2A}(\boldsymbol{\theta} - \mathbf{X}\boldsymbol{\beta})^T(\boldsymbol{\theta} - \mathbf{X}\boldsymbol{\beta})]\\
   &\times  & (\bar D + A)^{-\frac 12} .
 \end{eqnarray*}
From the above, the full conditionals are
\begin{itemize}
    \item [(A)] $\boldsymbol{\theta}| \boldsymbol{\beta}, A,\mathbf{y} \sim N(\boldsymbol{\delta}, \boldsymbol{\Phi}),$
    \item[(B)] $\boldsymbol{\beta}| \boldsymbol{\theta}, A,\mathbf{y}  \sim N((\mathbf{X}^T\mathbf{X})^{-1}\mathbf{X}^T\boldsymbol{\theta}, A(\mathbf{X}^T\mathbf{X})^{-1})$,
    \item[(C)] $\pi(A|\boldsymbol{\theta},\boldsymbol{\beta},\mathbf{y} ) \propto  (\bar D + A)^{-\frac 12} A^{-\frac m2} \exp[-\frac 1{2A}(\boldsymbol{\theta} - \mathbf{X}\boldsymbol{\beta})^T(\boldsymbol{\theta} - \mathbf{X}\boldsymbol{\beta})]$,
\end{itemize}
where
$$
\boldsymbol{\Phi} = (\mathbf{D}^{-1} + A^{-1}\mathbf{I}_m)^{-1}, ~ \boldsymbol{\delta} = \boldsymbol{\Phi}(\mathbf{D}^{-1}\mathbf{y} + A^{-1}\mathbf{X}\boldsymbol{\beta}),
$$
To sample from the conditional pdf of $A$ we use the rejection sampling. We use the inverse gamma distribution with shape parameter $(m-1)/2$ and rate parameter $(\boldsymbol{\theta} - \mathbf{X}\boldsymbol{\beta})^T(\boldsymbol{\theta} - \mathbf{X}\boldsymbol{\beta})/2$ as the proposal distribution. Alternatively, one can also use the Metropolis-Hastings algorithm within Gibbs with the same proposal as the transition density for $A$.

\clearpage

 \section{Techniques for comparing mass functions and sets}
 \label{Sec:Compare}

Note that the rows of $\mathbf{R}_{Bayes}$ are indexed by the ranks $\{ 1, \ldots, m \}$, consequently,  for any $j$, the entries of the $j$-th column $\{ \pi_{i, j, \mathbf{Y}} = \mathbf{R}_{Bayes, i, j}: i = 1, \ldots, m \} $ denote the relative weights on the ranks for the $j$-th small area. An estimate of the rank of the $j$-th small area, based on $\mathbf{R}_{Bayes, j}$ is the posterior mean $\hat{\mathbf{r}}_{j} = \sum_{i =1}^{m} i \pi_{i, j, \mathbf{Y}}$. We quantify the uncertainty of $\mathbf{R}_{Bayes, j}$ in capturing the true or estimated rank using the posterior \textit{mean absolute deviation} defined as 
$AMD_{j} = \sum_{i =1}^{m} | i - \check{\mathbf{r}}_{j}| \pi_{i, j, \mathbf{Y}}$ if $\check{\mathbf{r}}_{j}$ is known (as in simulations or in the presence of ground truth), and $\widehat{AMD}_{j} = \sum_{i =1}^{m} | i - \hat{\mathbf{r}}_{j}| \pi_{i, j, \mathbf{Y}}$ otherwise. Overall uncertainties are measured using $\overline{AMD} = m^{-1} \sum_{j = 1}^{m} AMD_{j}$ and 
$\widetilde{AMD} = m^{-1} \sum_{j = 1}^{m} \widehat{AMD}_{j}$. 

To quantify the spread of weights associated with ranks as in $\mathbf{R}_{Bayes, j}$, we use \textit{effective cardinality}, defined as $EC_{j} = \exp[ - \sum_{i =1}^{m} \pi_{i, j, \mathbf{Y}} \log (\pi_{i, j, \mathbf{Y}})]$, the exponential of the entropy. Overall spread is quantified with $\overline{EC} = m^{-1} \sum_{j=1}^{m} EC_{j}$. Smaller values of $\overline{EC}$ imply the distribution is more concentrated, hence more informative for predictions and inferences. If $\pi_{i, j, \mathbf{Y}} \equiv m^{-1}$ we have $EC_{j} = m$, the cardinality of the set of ranks. See Section~\ref{appendix-ellipse-size} for additional details including linkage of this with known metrics of mass concentration.
We use two metrics to compare the sizes of $\mathcal{S}_{KWW, \alpha}$ (or $\mathcal{R}_{KWW}$) and the partially ordered credible sets $\mathcal{S}_{UB, C, \alpha}$, $\mathcal{S}_{UB, E, \alpha}$, $\mathcal{S}_{HB, C, \alpha}$, and $\mathcal{S}_{HB, E, \alpha}$, (or $\mathcal{R}_{UB, C}$, $\mathcal{R}_{UB, E}$, $\mathcal{R}_{HB, C}$, $\mathcal{R}_{HB, E}$). The first measure is the \textit{volume} of these sets {(for $\boldsymbol{\theta}$)}, whose formula is well-known and also given in supplementary Section~\ref{appendix-ellipse-size}. The other metric is a measure of average \textit{length}, which for orthotopes is $m^{-1} \sum_{j} (U_{j} - L_{j})$. Since no equivalent metric exists for hyperellipses, consequently we propose one using lower-dimensional manifolds in Section~\ref{appendix-ellipse-size}.

\clearpage

\section{Size measures of elliptical sets} \label{appendix-ellipse-size}

As an alternative to an orthotope (or hyperrectangular) credible set, where each component of $\boldsymbol{\theta}$ varies freely, we have considered HPD sets which are approximately elliptical for our Bayesian models. Two size measures that appear to be appropriate for set estimators are their Lebesgue volumes and  average lengths of the ``sides'' determining the set. These two measures are easily found for an orthotope. We  compute representative values of these size measures of the elliptical sets.
	
\noindent  To determine reasonable size measures for elliptical regions, we proceed with the set 
	\begin{equation}
	\label{ellipse-A}
	E = \{\mathbf{x} : \mathbf{x}^T \boldsymbol{\Psi} \mathbf{x} \le c\},
	\end{equation}
	where $\boldsymbol{\Psi}$ is an $m\times m$, p.d. matrix, and $c>0$, determine the region. Since the center of the ellipse has no role in the size measures, we take it zero. Let $V_E(\boldsymbol{\Psi},m,c)$ denote the Lebesgue volume of the set $E$. It can be easily calculated that
	\begin{equation}
	\label{volEllip-A}
	V_E(\boldsymbol{\Psi},m,c) = \frac{\pi^{m/2}}{\Gamma ((m+2)/2)} \times \frac{c^{m/2}}{\mbox{det}(\boldsymbol{\Psi})^{1/2}},
	\end{equation}
	where $\Gamma(\cdot)$ is the gamma function and $\mbox{det}(\boldsymbol{\Psi})$ is the determinant of $\boldsymbol{\Psi}$.
	
\noindent A standard length measure of an elliptical region is not available. We propose below a reasonable expression for it. For an $m$-dimensional rectangular region, its volume is the product of the volume of the $(m-1)$-dimensional of the cross-section of the set defined by the intersecting hyperplane, say, $\theta_1 = d$, and the length $(U_1-L_1)$ of original rectangular set along the $\theta_1$ direction (where $L_1 < d < U_1$). Thus, the length of the set along the $x_1$ direction is the ratio of the volume of the original orthotope to the volume of the lower-dimensional cross-section orthotope. 
	
\noindent We will use the above discussion for the orthotope region as an analogy to come up with a suitable ``length'' for the ellipse along the $x_1$ direction. For the ellipse, we take for any $x_1$ and ${\mathbf{x}}_{(2)} =(x_2,\cdots, x_m)^T$ with $(x_1, {\mathbf{x}}_{(2)}^T)^T$ inside the set, we define a central cross-section by the hyperplane $x_1 = -\boldsymbol{\Psi}_{12}{\mathbf{x}_{(2)}}/\boldsymbol{\Psi}_{11}$, where we partition the matrix $\boldsymbol{\Psi}$ as $((\boldsymbol{\Psi}_{ij}))$, $i,j=1,2$, in conformity with the partitioning of $\mathbf{x}$ above. We also define the p.d. matrix $\boldsymbol{\Psi}_{22.1} = \boldsymbol{\Psi}_{22} - \boldsymbol{\Psi}_{21}\boldsymbol{\Psi}_{11}^{-1}\boldsymbol{\Psi}_{12}$. The corresponding cross-section is $E_C =\{\mathbf{x}_{(2)}^T \boldsymbol{\Psi}_{22.1}\mathbf{x}_{(2)} \le c\}.$ From equation (\ref{volEllip-A}), it similarly follows that 
	\begin{equation} \label{vol-cross-A}
	V_{E_C}(\boldsymbol{\Psi}_{22.1},m-1,c) = \frac{\pi^{(m-1)/2}}{\Gamma ((m+1)/2)} \times \frac{c^{(m-1)/2}}{\mbox{det}(\boldsymbol{\Psi}_{22.1})^{1/2}} .   
	\end{equation}
	
\noindent A reasonably representative length for the $x_1$ side of the ellipse is the ratio of the volume in (\ref{volEllip-A}) to the volume in (\ref{vol-cross-A}). Simplification leads to an expression. From this, after simplification, in general,  we have along the $x_i$ direction the representative length, $L_{R,i}$ (say), is
	\begin{equation}
	\label{LRi}
	L_{R,i} = B\left(\frac 12, \frac{m+1}2\right) \times \left\{\frac{\sqrt{c}}{a_i} \right\},~ i=1,\cdots, m,
	\end{equation}
where $a_i^2$ is the $i$th diagonal of $\boldsymbol{\Psi}$ and $B(a,b)$ is the regular beta function.
	
\noindent The above proposal of the length of a side of an elliptical region is only heuristic. Indeed, it is a reasonable requirement that these lengths should be ``calibrated'' by a suitable scale so that for the modified orthotope based on the  lengths of the ``scaled sides'' will match the volume of the original ellipse. Based on this consideration, the modified lengths $L_{M,i}$ are given by 
	\begin{equation}
	\label{lin-measure-ellip}
	L_{M,i} = \{V_E(\boldsymbol{\Psi},m,c)/\prod_{i=1}^m L_{R,i}\}^{1/m} L_{R,i} . 
	\end{equation}
$L_E\stackrel{def} =\frac 1m \sum_{i=1}^mL_{M,i}$,  a ``linear'' measure of the elliptical set, may be compared to $L_C$.

\subsection{Comparison via entropy} \label{entropy}	

For any discrete distribution  on values $v_1, \cdots, v_n$ with respective weights $f_1, \cdots, f_n$, define the \textit{effective cardinality} $C_f = \exp[E_f]$, where $E_f = -\sum f_j \log(f_j)$ is the entropy. Note that smaller values of $E_f$ and consequently $C_{f}$ implies the distribution is more concentrated, hence more informative for predictions and inferences. Note that for the uniform probability mass function $f_j=1/n, j=1,\cdots, n,$ we have $C_f = n$, which matches the notion of cardinality of a finite set.
Also note that the effective cardinality  is a special case of Hill's number of order $q$ defined by
$^qD = \{\sum_{i=1}^n f_i^q\}^{1/(1-q)}$,
which is used to measure the diversity of an ecological system. 
Frequently suggested values of $q$ are $0,1,2$ and infinity. 
For $q=1$ and $q=\infty$, the index is defined taking the limit. The index $^1D$ is the effective cardinality defined earlier.  Also, $^\infty D = 1/f_{(n)}$, provided that $f_{(n)}$ is a unique maximum of the $f_i$'s, and in this case, since $f_{(n)} > 1/n$,  $^\infty D < n$; the upper bound to $^\infty D$ is reached by a uniform distribution over the $n$ elements. For $q=2$, the index $^2D = (\sum_{i=1}^n f_i^2)^{-1} \le n$, with the equality again prevailing for a uniform distribution; cf. \cite{hill1973diversity}.

 \begin{Proposition} \label{Prop:Cardinality}
 For the Bayesian empirical credible sets ${\cal F}_{KWW^{(B)}, \alpha , \mathbf{y}}$, ${\cal F}_{UB, \alpha , \mathbf{y}}$ and ${\cal F}_{HB, \alpha, \mathbf{y}}$ the effective cardinalities are less than or equal to the number of elements in the respective sets. For the confidence set of ranks  ${\cal F}_{KWW, \alpha , \mathbf{y}}$ obtained by the frequentist KWW approach, the effective cardinality is same as its cardinality.
\end{Proposition}
 
The  proof of Proposition~\ref{Prop:Cardinality} easily follows using Jensen's inequality and by the nonnegativity of the Kullback-Leibler divergence. We omit the details.

\clearpage

\section{Theoretical details}
\label{Sec:Theory_Supp}

We assume the existence of a probability space $(\Omega, \mathcal{B}, \mathbb{P})$ throughout, where $\Omega$ bears the interpretation of the sample space, and $\omega \in \Omega$ is any element thereof. In the following, we consider without loss of generality $m_{n} \equiv n$ for simplicity in presentation. Without loss of generality, fix $n_{1} < n_{2}$. For a  matrix $\mathbf{A}$, let $|\mathbf{A}|$ denote its Frobenius norm,  $\text{tr} (\mathbf{A})$ denote its trace, and $\text{det} (\mathbf{A})$ denote its determinant. Also, for any matrix $\mathbf{A} \in \mathbb{R}^{n_{2} \times n_{2}}$, the notation $\mathbf{A}^{(n_{1})}$ stands for its leading $n_{1} \times n_{1}$ submatrix. The spectral representation for a symmetric nonnegative definite matrix  $\boldsymbol{\Sigma}_{j}$ is given by $\boldsymbol{\Sigma}_{j} = \mathbf{P}_{j} \boldsymbol{\Lambda}_{j} \mathbf{P}_{j}^{T}$ for $j = 1, 2$, where the eigenvalues in $\boldsymbol{\Lambda}_{j} $ are in decreasing order. Unless stated otherwise, the $D_{n, i}$ is assumed to be bounded above and below for each $i$ and each $n$. The standard technical framework applicable to all our results is given below.

\noindent\textbf{Assumption-A:} (Basic framework)
For every $n$ sufficiently large,  the matrix \\ $\mathbf{X}_{n} = (\mathbf{x}_{n, 1}, \ldots, \mathbf{x}_{n, m_{n}} )^{T} \in \mathbb{R}^{m_{n} \times p}$ is full column rank for each sufficiently large $n$, and the maximum diagonal entry of the projection matrix on the column space of $\mathbf{X}_{n}$ is $O( n^{-1})$. We assume that there exist true but unknown parameters $({\boldsymbol{\beta}}_{\infty}, A_{\infty}) \in \mathbb{R}^{p} \times (0, \infty)$.

The $\Delta$-FH framework and our technical conditions of \textbf{Assumption-A} ensure posterior concentration of the posterior distribution of $({\boldsymbol{\beta}}_{n}, A_{n})$, see \cite{ref:AoS952145_GGS, ref:AoS00500_GhosalGhoshvanderVaart_PosteriorConcentration, ref:AoS07192_GhosalvanderVaart_PosteriorConcentration} for technical details on how such results are established.  In particular, for a fixed positive $\delta_{*}< A_{\infty}/8$, the posterior probability $\mathbb{P} [ A_{n} < \delta_{*} | \mathbf{Y}_{n}]$ is exponentially small, consequently, we redefine $A_{n} = \max\{\delta_{*}, A_{n} \}$ to avoid unnecessary technicalities. The assumption of a lower bound for variance components (or their estimators) for mathematical results is quite common, see \cite{jiang2007linear, jiang2010large} for example.

\subsection{Posterior distribution details}
In the Fay-Herriott model, the joint distribution of the observed data $\mathbf{Y} = (Y_{1}, \ldots, Y_{n})^{T}$ and the unknown 
$\boldsymbol{\theta} = (\theta_{1}, \ldots, \theta_{n})^{T}$ can be written as (conditional on the parameters in the Bayesian viewpoint)
\begin{align*}
\left( \begin{array}{l}
\mathbf{Y}_{n} \\
\boldsymbol{\theta}_{n} \end{array} \right) & \sim  
N_{2n} \left( 
\left( \begin{array}{l}
\mathbf{X}_{n} {\boldsymbol{\beta}}_{n} \\
\mathbf{X}_{n} {\boldsymbol{\beta}}_{n} \end{array} \right), 
\left( \begin{array}{ll}
A_{n} \mathbb{I}_{n} + \mathbf{D} & A_{n} \mathbb{I}_{n} \\
A_{n} \mathbb{I}_{n}  & A_{n} \mathbb{I}_{n} \end{array}\right) \right)\\
& =  N_{2n} \left( 
\left( \begin{array}{l}
\boldsymbol{\mu}_{n} \\
\boldsymbol{\mu}_{n} \end{array} \right), 
\left( \begin{array}{ll}
{\mathbf{V}}_{11} & {\mathbf{V}}_{12}  \\
{\mathbf{V}}_{21} & {\mathbf{V}}_{22}\end{array}\right) \right).
\end{align*}

The unobserved $\boldsymbol{\theta}_{n}$ has the following conditional distribution given $\mathbf{Y}_{n}$ and the parameters:
\begin{align}
    (\boldsymbol{\theta}_{n} | \mathbf{Y}_{n}, {\boldsymbol{\beta}}_{n}, A_{n}) & \sim  N_{n} ( \boldsymbol{\nu}_{n}, \boldsymbol{\Gamma}_{n} ), \text{ where} \label{eq:posterior}\\
\boldsymbol{\nu}_{n} & =  \boldsymbol{\mu}_{n} 
+ {\mathbf{V}}_{21} {\mathbf{V}}_{11}^{-1} (\mathbf{Y}_{n} - \boldsymbol{\mu}_{n}),  \nonumber\\
\boldsymbol{\Gamma}_{n}  & =  {\mathbf{V}}_{22} - {\mathbf{V}}_{21}{\mathbf{V}}_{11}^{-1} {\mathbf{V}}_{12}. \nonumber
\end{align}

Let $B_{n, i}= D_{n, i}/(D_{n, i} + A_{n}), \ i= 1, \ldots, m_{n}$(=$n$). Then in the Fay-Herriot model, 
${\mathbf{V}}_{21}{\mathbf{V}}_{11}^{-1} =  \mathbb{I}_{n} - \mathbf{B}_{n}$ where $\mathbf{B}_{n} = {\rm diag} (  B_{n, 1}, \ldots, B_{n, m_{n}})$. 

Thus, 
\begin{align*}
\boldsymbol{\nu}_{n} & =  \boldsymbol{\mu}_{n} + (\mathbb{I}_{n} - \mathbf{B}_{n})(\mathbf{Y}_{n} - \boldsymbol{\mu}_{n}) 
= \mathbf{B}_{n} \boldsymbol{\mu}_{n} + (\mathbb{I}_{n} - \mathbf{B}_{n})\mathbf{Y}_{n}, \\
\boldsymbol{\Gamma}_{n} & =  A_{n} \mathbb{I}_{n} - A_{n} (\mathbb{I}_{n} - \mathbf{B_{n}}) = A_{n} \mathbf{B}_{n}.
\end{align*}

\subsection{Proofs of theorems}

We have omitted the considerable parts of inevitable but routine algebra in the proofs below, and discuss only the main ideas. In particular,  computations involving multivariate Gaussian distribution to show that certain remainder terms are negligible are generally omitted below. 

\begin{Proof}[Proof of Theorem~\ref{Thm:1}]
    Associated with a permutation $\boldsymbol{\sigma}_{n}$ of $\mathbb{N}_{n}$, define the open (and hence Borel) cone 
\begin{align*}
C (\boldsymbol{\sigma}_{n}) & = \bigl\{ \mathbf{x} \in \mathbb{R}^{n}: x_{\sigma_{1}} < x_{\sigma_{2}} < \ldots < x_{\sigma_{n}}\bigr\}.
\end{align*}
Note that for any $\mathbf{x} \in  C (\boldsymbol{\sigma}_{n})$,  $c \mathbf{x} \in C (\boldsymbol{\sigma}_{n})$ for all $c > 0$. 
It can be immediately seen that the event $\mathbf{R}_{n}= \boldsymbol{\sigma}_{n}$ is the same as $n^{-1/2}  \boldsymbol{\theta}_{n} \in C (\boldsymbol{\sigma}_{n})$, consequently we study  
\begin{align*}
    \mathbb{P} [n^{-1/2} \boldsymbol{\theta}_{n} \in C (\boldsymbol{\sigma}_{n}) | \mathbf{Y}_{n}, \boldsymbol{\beta}_{n}, A_{n}].
\end{align*}

Now it is immediate that \eqref{eq:Thm1}  is bounded by the total variation distance between $\boldsymbol{\theta}_{n_{1}}$ and $\boldsymbol{\theta}_{n_{2}}^{(n_{1})}$, which in turn is bounded above $\sqrt{2}$ times the Hellinger distance between them \cite{ref:Arxiv1810.08693v7_TotalVariation_Devroye}.

Consider the case where $n_{1}$ and $n_{2}$ are the same order, the situation where $n_{1} = o (n_{2})$ presents no additional technical difficulties and can be handled similarly. 
Outside a null set, we have that \eqref{eq:Thm1}  is bounded by 
\begin{align*}
   & C_{n_{1}, n_{2}}  (\boldsymbol\beta_{n}, A_{n}) \\
     = & \  2^{1/2} \biggl[ 1 - 
    \exp \biggl\{ - {\frac{1}{4}} \bigl( \boldsymbol{\nu}_{n_{1}} - \boldsymbol{\nu}_{n_{2}}^{(n_{1})}\bigr)^{T} \bigl(\boldsymbol{\Gamma}_{n_{1}} + \boldsymbol{\Gamma}_{n_{2}}^{(n_{1})} \bigr)^{-1}
    \bigl( \boldsymbol{\nu}_{n_{1}} - \boldsymbol{\nu}_{n_{2}}^{(n_{1})}\bigr)
    \\ & \hspace{0.5cm}
    +   {\frac{1}{4 n_{1}}}\log\text{det} (\boldsymbol{\Gamma}_{n_{1}})
    +   {\frac{1}{4 n_{1}}}\log\text{det} (\boldsymbol{\Gamma}_{n_{2}}^{(n_{1})})
    -   {\frac{1}{2 n_{1}}}\log\text{det} \Bigl( {\frac {\boldsymbol{\Gamma}_{n_{1}} + \boldsymbol{\Gamma}_{n_{2}}^{(n_{1})}}{2}} \Bigr)
    \biggr\} \biggr]^{1/2}\\
     = & \ 2^{1/2} \bigl(1 - \exp \{- T\}  \bigr)^{1/2} \hspace{1cm} \text{(say)} \\
     \leq & \ 2^{1/2} T^{1/2}.
\end{align*}

To show that this is $O(n_{1}^{-1/2})$ almost surely we need several additional steps that are algebraic and quite routine in nature, and we omit much of the details. Define $T_{n} = (A_{n} - A_{\infty})/A_{\infty}$ and $T_{n, i} = (A_{n} - A_{\infty})/(D_{n, i} + A_{\infty})$, and notice that under standard posterior convergence results, both of these are $O(n^{-1/2})$ almost surely. Define 
$B_{\infty, i}= D_{n, i}/(D_{n, i} + A_{\infty}), \ i= 1, \ldots, m_{n}$, $\mathbf{B}_{\infty} = {\rm diag} (  B_{\infty, 1}, \ldots, B_{\infty, m_{n}})$ and $\boldsymbol{\Gamma}_{\infty} = A_{\infty} \mathbf{B}_{\infty}$. Then it is possible to show that 
$\boldsymbol{\Gamma}_{\infty}^{-1} \boldsymbol{\Gamma}_{n} = \mathbb{I}_{n} + {\rm diag}(T_{n} + T_{n, i}) + {\rm diag} (R_{n, i})$, 
where $| {\rm diag} (R_{n, i}) | = O(n^{-1})$ almost surely. We now leverage these calculations, our assumptions and the known properties of multivariate Gaussian distributions, and the fact that only a countable number of null sets are collected, to arrive at the result.
The first term in the above expression for $C_{n_{1}, n_{2}}  (\boldsymbol\beta_{n}, A_{n})$ is easily seen to be $O(n_{1}^{-1})$ almost surely. In the other terms, there are cancellations of the terms involving $\boldsymbol{\Gamma}_{\infty}$ and terms up to $O_{P}(n_{1}^{-1/2})$,  leaving only higher order terms in $T$.

\end{Proof}

\noindent\textbf{Assumption-B:} We assume that  the triangular sequence $\{ D_{n, i}: \ i = 1, \ldots, n; \ n = 1, 2, \ldots \}$ is a 
nested sequence. That is, for any $n_{1} < n_{2}$, $D_{n_{2}, i} = D_{n_{1}, i}$ for $i = 1, \ldots, n_{1}$. In addition, we assume the following limit exists and is finite:
\begin{align}
    \lim_{n \to \infty} n^{-1} \sum_{i = 1}^{n} {\frac{D_{n, i}}{A_{\infty} + D_{n, i}}} = L \in (0, 1).
    \label{eq:BoundedVar}
\end{align}

Since we make the routine assumption that the $D_{n, i}$'s are bounded above and below, say by $D_{max} < \infty$ and $D_{min} > 0$ respectively,  the existence of the limit in \eqref{eq:BoundedVar} is not an additional condition. Note that the sequence \{ $n^{-1} \sum_{i = 1}^{n} {\frac{D_{n, i}}{A_{\infty} + D_{n, i}}} \}$ is bounded, and hence by Bolzano-Weierstrass Theorem has a convergent subsequence. The limit in \eqref{eq:BoundedVar} by considering the $n_{i}$'s along this subsequence.

\begin{Proof}[Proof of Theorem~\ref{Thm:2}]

We will show that the Gromov-Wasserstein distance between the distributions $\mathbb{G}_{n_{1}} = [n_{1}^{-1/2}\boldsymbol{\theta}_{n_{1}}| \mathbf{Y}_{n_{1}}] = N_{n_{1}} \bigl( n_{1}^{-1/2}\boldsymbol\nu_{1}, n_{1}^{-1}\boldsymbol{\Gamma}_{n_{1}} \bigr)$ and $\mathbb{G}_{n_{2}} = [n_{2}^{-1/2} \boldsymbol{\theta}_{n_{2}}| \mathbf{Y}_{n_{2}}] =  N_{n_{2}} \bigl( n_{2}^{-1/2}\boldsymbol\nu_{2}, n_{2}^{-1}\boldsymbol{\Gamma}_{n_{2}} \bigr)$ converge to zero as $\min \{ n_{1}, n_{2} \} \to \infty$, thus establishing a Cauchy convergence type result on measure spaces. Note that the distributions $\mathbb{G}_{n_{1}}$ and $\mathbb{G}_{n_{2}}$ are random and supported on different spaces.

Note that the actual Gromov-Wasserstein distance between $\mathbb{G}_{n_{1}}$ and $\mathbb{G}_{n_{2}}$ is not available in a closed form. 
However, using \cite{ref:JAppliedProbability221178}, we obtain the following upper bound for the square of the Gromov-Wasserstein distance between $\mathbb{G}_{n_{1}}$ and $\mathbb{G}_{n_{2}}$:
\begin{align}
    U (\mathbb{G}_{n_{1}}, \mathbb{G}_{n_{2}})^{2} & = 4 \bigl| \text{tr} (n_{2}^{-1}\boldsymbol{\Gamma}_{n_{2}}) - \text{tr} (n_{1}^{-1}\boldsymbol{\Gamma}_{n_{1}})\bigr|^{2} 
              + 8 | n_{2}^{-1}\boldsymbol{\Gamma}_{n_{2}}^{(n_{1})} - n_{1}^{-1}\boldsymbol{\Gamma}_{n_{1}} |^{2} \nonumber \\
           & \hspace{3.5cm} + 8 \bigl( | n_{2}^{-1}\boldsymbol{\Gamma}_{n_{2}}|^{2} - | n_{2}^{-1}\boldsymbol{\Gamma}_{n_{2}}^{(n_{1})} |^{2}  \bigr)
           \label{eq:Upper_GW}\\
           & = T_{1} + T_{2} + T_{3}, \ \ \textrm{(say).} \nonumber
\end{align}

We will show that each of $T_{1}, T_{2}, T_{3}$ has the correct rate outside a null set, which will establish the result. As earlier, we skip much of the routine algebraic details below.

For $T_{1}$, notice that outside a null set, we have 
\begin{align*}
    T_{1} & \leq 4 \left| n_{2}^{-1} A_{n_{2}} \sum_{i = 1}^{n_{2}} {\frac{ D_{n_{2}, i}}{A_{\infty} + D_{n_{2}, i}}} - n_{1}^{-1} A_{n_{1}} \sum_{i = 1}^{n_{1}} {\frac{ D_{n_{1}, i}}{A_{\infty} + D_{n_{1}, i}}}\right| + n_{1}^{-1}T_{1,1}, 
\end{align*}
where $T_{1, 1}$ is a $O (1)$ random variable almost surely, using arguments similar to those used in the proof of Theorem~\ref{Thm:1}. Now we obtain the result by using the conditions laid out in \textbf{Assumption-B}. 

We decompose $T_{2}$ as 
\begin{align*}
    T_{2} & = 8 | n_{2}^{-1}\boldsymbol{\Gamma}_{n_{2}}^{(n_{1})} - n_{1}^{-1}\boldsymbol{\Gamma}_{n_{1}} |^{2} \\
    & \leq 8 | n_{2}^{-1}\boldsymbol{\Gamma}_{n_{2}}^{(n_{1})} - n_{1}^{-1}\boldsymbol{\Gamma}_{n_{2}}^{(n_{1})} |^{2}
    + 8 | n_{1}^{-1}\boldsymbol{\Gamma}_{n_{2}}^{(n_{1})} - n_{1}^{-1}\boldsymbol{\Gamma}_{n_{1}} |^{2} \\
    & = T_{2, 1} + T_{2, 2}.
\end{align*}
Now note that for a $O (1)$ random variable $T_{2, 3}$ almost surely
\begin{align*}
    | T_{2, 1} |^{2} & = (1 - n_{1}/n_{2})^{2} n_{1}^{-2} A_{n_{2}}^{2} \sum_{i = 1}^{n_{1}} {\frac{ D_{n_{1}, i}^{2}}{(A_{\infty} + D_{n_{1}, i})^{2}}} + n_{1}^{-1}T_{2,3} \\
    & \leq n_{1}^{-1} A_{n_{2}}^{2} L + n_{1}^{-1}T_{2,3},
 \end{align*}
thus establishing  that this term is $O (n_{1}^{-1})$ almost surely. A simple algebraic computation with steps similar to those used earlier should show that $| T_{2, 1} |^{2}  = O (n_{1}^{-1})$ almost surely also.

It is easy to see that outside a null set for a $O (1)$ random variable $T_{3, 1}$
\begin{align*}
    T_{3} & = 8 n_{2}^{-2} A_{n_{2}}^{2}  \sum_{i = n_{1} + 1}^{n_{2}} {\frac{ D_{n_{2}, i}^{2}}{(A_{n_{2}} + D_{n_{2}, i})^{2}}} \\
    & \leq  8 n_{2}^{-1} A_{n_{2}}^{2} L + + n_{2}^{-1}T_{3,1},
\end{align*}
which establishes the required result.
\end{Proof}

\end{document}